\documentclass[12pt,twoside,a4paper]{article}
\usepackage{amsmath,amssymb,latexsym,theorem,bbm,natbib,epsfig,color,subfigure}
\usepackage{multirow,graphicx,array,rotating}  
\newfont{\msa}{msam10 scaled\magstep1}
\newfont{\ssmsa}{msam9}

\def\crps{\mathop{\hbox{\rm CRPS}}}
\def\twcrps{\mathop{\hbox{\rm twCRPS}}}
\def\twcrpss{\mathop{\hbox{\rm twCRPSS}}}

\numberwithin{equation}{section}

\title{Log-normal distribution based EMOS models for
  probabilistic wind speed forecasting}

\author{{\sc S\'andor Baran$^{1}$} and {\sc Sebastian Lerch$^{2}$}\\
         $^1$Faculty of Informatics, University of Debrecen\\
         Kassai \'ut 26, H-4028 Debrecen, Hungary \\
         $^2$ Institute of Applied Mathematics, Heidelberg University\\
        Im Neuenheimer Feld 294, D-69120 Heidelberg, Germany}

\date{}

\begin{document}
\pagestyle{myheadings}

\maketitle

\begin{abstract}
Ensembles of forecasts are obtained from multiple runs of numerical
weather forecasting models with different initial conditions and
typically employed to account for forecast uncertainties. However, biases and
dispersion errors often occur in forecast ensembles, they are
usually under-dispersive and uncalibrated and require statistical
post-processing.  We present an Ensemble Model Output Statistics (EMOS)
method for 
calibration of wind speed forecasts based on the log-normal (LN)
distribution, and we also show a regime-switching extension of the
model  which combines the previously studied truncated normal (TN)
distribution with the LN. 

Both presented models are applied to wind speed forecasts of the
eight-member University of Washington mesoscale ensemble, of the
fifty-member ECMWF ensemble and of the eleven-member ALADIN-HUNEPS
ensemble of the Hungarian Meteorological Service, and their predictive
performances are compared to those of the TN and general extreme value
(GEV) distribution based EMOS methods and to the TN-GEV mixture model.
The results indicate improved calibration of
probabilistic and accuracy of point forecasts in comparison to the raw
ensemble and to climatological forecasts. Further, the TN-LN
mixture model outperforms the traditional TN method and its predictive
performance is able to keep up with the models utilizing the GEV
distribution without assigning mass to negative values.

\end{abstract}

\section{Introduction}
  \label{sec:sec1}

Accurate and reliable forecasting of wind speed is of 
importance in various field of economy, e.g., agriculture,
transportation, energy production. Forecasts are usually based on
current observational data and mathematical models
describing the dynamical and physical behaviour of the atmosphere. 
These models consist of sets of coupled hydro-thermodynamic non-linear
partial differential equations which have only numerical solutions and
highly depend on initial conditions. To reduce the uncertainties
coming either from the lack of reliable initial conditions or from the
numerical weather prediction process itself, a possible solution is to
run the models with different initial conditions resulting in an
ensemble of forecasts  \citep{leith}. Since its first operational
implementation \citep{btmp,tk} the ensemble method has become a widely used
technique all over the world. One of the leading organizations issuing
ensemble forecasts is the  European
Centre for Medium-Range Weather Forecasts \citep{ecmwf}, but 
all major national meteorological services have their own ensemble
prediction systems (EPS), e.g., the COSMO-DE EPS of the German
Meteorological Service \citep[DWD; ][]{gtpb,btg} or the
PEARP EPS of M\'eteo France \citep{dljn}. Besides calculating the
classical point 
forecasts (e.g. ensemble mean or ensemble median) using a forecast
ensemble one can also estimate the distribution of a future weather
variable which allows probabilistic forecasting \citep{grsc}. However,
the forecast ensemble is usually under-dispersive and as a
consequence, uncalibrated. This phenomenon has been observed with
several operational ensemble prediction systems \citep[see,
e.g.,][]{bhtp}. A possible solution to account for this deficiency is
statistical post-processing.

From the various modern post-processing techniques 
(for an overview see, e.g., \citet{wfk,gneiting14}) 
probably the most
widely used methods are the Bayesian model averaging (BMA) introduced by
\citet{rgbp} and the ensemble model output statistics (EMOS) or
non-homogeneous regression technique, suggested
by \citet{grwg}, as they are   
implemented in {\tt ensembleBMA} \citep{frgs,frgsb} and {\tt ensembleMOS}
packages of {\tt R}. Both approaches provide estimates of the densities
of the predictable weather quantities and once a predictive density is
given, a point forecast can be easily determined (e.g., mean or median value). 

The BMA predictive probability density function (PDF) of a future
weather quantity is the weighted sum of individual PDFs corresponding
to the ensemble members. An individual PDF can be interpreted as the
conditional PDF of the future weather quantity provided the considered
forecast is the best one and the weights are based on the
relative performance of the ensemble members during a given training
period. In the case of wind speed \citet{sgr10} suggest the use of
a gamma mixture while \citet{bar} considers BMA component PDFs 
following a truncated normal (TN) rule. 

The EMOS approach uses a single parametric distribution as a
predictive PDF with parameters depending on the ensemble members. 
The unknown parameters specifying this dependence are estimated using
forecasts and validating observations from a rolling training period,
which allows automatic adjustments of the statistical model to any
changes of the EPS system (for instance seasonal variations or EPS
model updates). For wind speed \citet{tg} suggest to use a
TN distribution, while \citet{lt} consider a generalized extreme value (GEV)
distributed predictive PDF. To ensure a more accurate prediction of
high wind speed values the authors also introduce a TN-GEV
regime-switching model where the use of the two distributions depend
on the value of the ensemble median: for large values a GEV, otherwise a
TN based EMOS model is applied.

In the present paper we develop an EMOS model where the predictive PDF
follows a log-normal (LN) distribution. Besides this, similar to
\citet{lt}, we propose a TN-LN regime-switching mixture model, where
an LN distribution is applied to high wind speed values and the
choice again depends on the ensemble median.
Compared to the GEV distribution approach of \citet{lt} the main
advantage of the LN model 
is its computational simplicity, which allows faster estimation of
model parameters. The predictive performance of the LN model and of
the TN-LN mixture model is tested on forecasts
of maximal wind speed of the eight-member University of 
Washington Mesoscale Ensemble \citep[UWME, see e.g.,][]{em05} and of the
ECMWF ensemble \citep{lp}, and on instantaneous wind speed forecasts
produced by the operational Limited Area Model Ensemble  
Prediction System of the Hungarian Meteorological Service
(HMS) called ALADIN-HUNEPS \citep{hagel, horanyi}. These three
ensemble prediction systems (EPS) differ both in the generation of
ensemble forecasts and in the predictable wind quantities. As
benchmarks in all case studies we investigate the goodness of fit of
the TN model of \citet{tg} and of the GEV and TN-GEV mixture models of
\citet{lt}.  

\section{Data}
  \label{sec:sec2}

\subsection{University of Washington Mesoscale Ensemble}
  \label{subs:subs2.1}
The eight members of the UWME are obtained from different runs of the
fifth generation Pennsylvania State  
University--National Center for Atmospheric Research mesoscale model 
(PSU-NCAR MM5) with initial conditions from different sources
\citep{grell}. The EPS covers the Pacific Northwest region of western
North America providing forecasts on a 12 
km grid. Our data base (identical to the one used in
\citet{mlt}) contains ensembles of 48 h forecasts 
and corresponding validating observations of 10 m maximal wind
speed (maximum of the hourly instantaneous wind speeds over the 
previous twelve hours, given in m/s, see e.g. \citet{sgr10}) for
152 stations in the Automated Surface
Observing Network \citep{asos} in the states of Washington, Oregon,
Idaho, California and Nevada in the United States. The forecasts are 
initialized at 0 UTC (5 pm local time when daylight saving time (DST) is in use
and 4 pm otherwise) and the generation of the ensemble ensures that
its members are not exchangeable. In the present study we investigate
only forecasts for calendar year 2008 with additional data from the
last month of 2007 used for parameter estimation. Standard quality control
procedures were applied to the data set
and after removing
days and locations with missing data 101 stations remain where
the number of days for which forecasts and validating observations are
available varies between 160 and 291.

\begin{figure}[t]
\begin{center}
\leavevmode
\includegraphics[width=\textwidth]{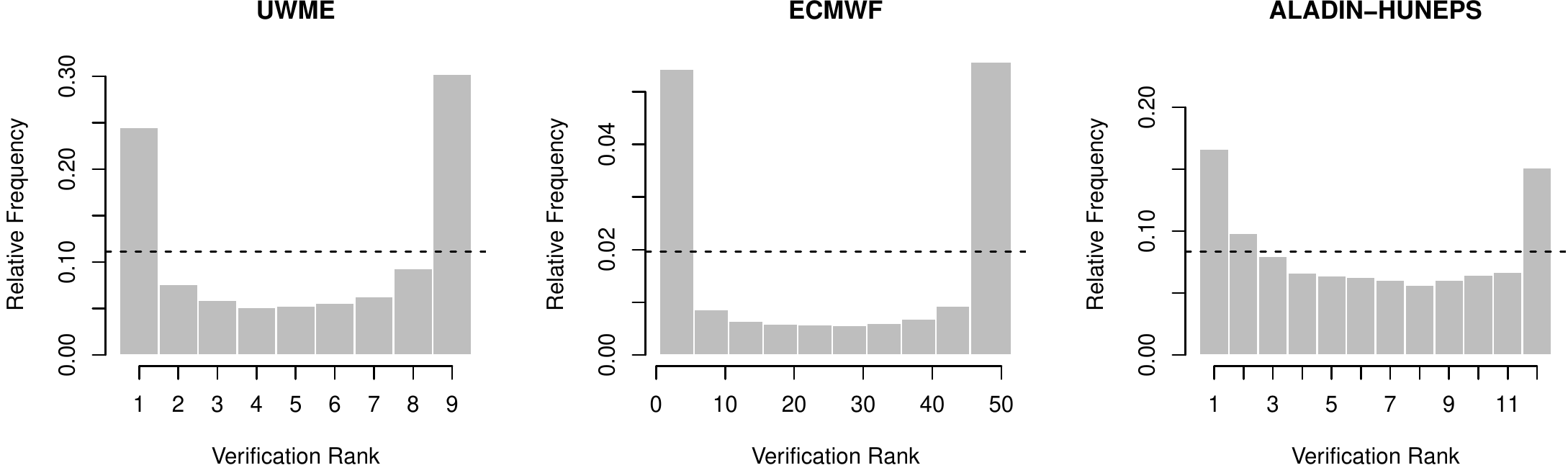}
\centerline{\hbox to 12cm{\qquad\small (a) \hfill  (b) \hfill (c)
    \hspace{-0.75cm}}} 
\end{center}
\caption{Verification rank histograms.  a) UWME
  for the calendar year 2008. b) ECMWF ensemble for the period
  May 1, 2010 -- April 30, 2011; c) ALADIN-HUNEPS
  ensemble for the period April 1, 2012 -- March 31, 2013.} 
\label{fig:fig1}
\end{figure}
Figure \ref{fig:fig1}a shows the verification rank histogram of the raw
ensemble, that is the histogram of ranks of validating
observations with respect to the corresponding ensemble
forecasts computed from the ranks at all locations and dates
considered \citep[see, e.g.,][Section 7.7.2]{wilks}. This histogram
is strongly U-shaped as in many cases the
ensemble members either underepredict or overpredict the validating
observations. The reliability index \ $\Delta = \sum_{i=1}^{c} \left|
  p_i - \frac{1}{c} \right|,$  \ 
where \ $c$ \ denotes the number of classes in the histogram, each of
which has expected relative frequency \ $1/c$, \ and \ $p_i$ \ denotes the
observed relative frequency in class \ $i$, 
can be used to quantify the deviation of the rank distribution from
uniformity \citep{delle}. For the UWME ensemble, \ $\Delta$ \ equals 
$0.6508$, and the ensemble range  
contains the observed maximal wind speed in only 
$45.24\,\%$ of the cases (the nominal value of this coverage
equals $7/9$, i.e $77.78 \,\%$). Hence, the ensemble is under-dispersive, thus
uncalibrated, and would require statistical post-processing to yield an
improved forecast probability density function. 

\subsection{ECMWF ensemble}
  \label{subs:subs2.2}

The global ensemble prediction system of the ECMWF consists of 50 exchangeable
ensemble members which are generated from random perturbations in
initial conditions and stochastic physics parametrization
\citep{mbp,lp}. Forecasts of near-surface (10 meter) wind speed for
lead times up to 10 days ahead are issued twice a day at 00 UTC and 12
UTC, with a horizontal resolution of about 33 km. Following
\citet{lt}, we focus on the ECMWF  
ensemble run initialized at 00 UTC (2 am local time when DST
operates and 1 am otherwise) and one day ahead
forecasts. Predictions of daily maximum wind speed are obtained as the
daily maximum of each ensemble member at each grid point location. 

The verification is performed over a set of 228 synoptic observation
stations over Germany. The validating observations are hourly
observations of 10-minute average wind speed measured over the 10
minutes before the hour.  Daily maximum wind speed observations are
given by the maximum over the 24 hours corresponding to the time
frame of the ensemble forecast. Ensemble forecasts at individual
station locations are obtained by bilinear interpolation of the
gridded model output.  Our results are based on a verification period
from May 1, 2010 to April 30, 2011, consisting of 83\,220 individual
forecast cases.  Additional data from February 1, 2010 to April 30,
2011 are used to allow for training periods of equal lengths for all
days in the verification period and for model selection purposes. 

The verification rank histogram of the ECMWF ensemble displayed in
Figure \ref{fig:fig1}b is even more U-shaped than that of the
UWME resulting in a reliability index of $1.1063$, while the ensemble
range contains the validating observation just in $43.40\,\%$ of all cases
(here the nominal value is $49/51$, that is 
$96.08\,\%$).  Again, the ensemble is under-dispersive 
and statistical calibration is required.

\subsection{ALADIN-HUNEPS ensemble}
  \label{subs:subs2.3}
The ALADIN-HUNEPS system of the HMS covers a large part of continental
Europe with a horizontal resolution 
of 12 km and is obtained with dynamical downscaling (by the ALADIN
limited area model) of the global
ARPEGE based PEARP system of M\'et\'eo France \citep{hkkr,dljn}. The
ensemble consists of 11 members, 10 initialized from perturbed initial
conditions and one control member from the unperturbed analysis,
implying that the ensemble contains groups of exchangeable
forecasts. 

The data base contains 11 member ensembles of 42 hour forecasts for 10
meter instantaneous wind speed (given in m/s) for 10 major cities in 
Hungary (Miskolc, Szombathely, Gy\H or, Budapest, Debrecen, Ny\'\i regyh\'aza,
Nagykanizsa, P\'ecs, Kecskem\'et, Szeged) produced by the
ALADIN-HUNEPS system of the HMS, together with the corresponding
validating observations for the one-year period between April 1, 2012
and March 31, 2013. 
The validating observations were scrutinized by basic 
quality control algorithms including, e.g., consistency checks.
The forecasts are initialized at 18 UTC (8 pm local time when DST
operates and 7 pm otherwise). The data set is fairly complete
since there are only six days when no forecasts are available. These dates are
excluded from the analysis.  

Similar to the previous two examples, the verification rank histogram of
the raw ALADIN-HUNEPS ensemble is far
from the desired uniform distribution (see Figure \ref{fig:fig1}c),
however, it shows a less 
under-dispersive character. The better fit of the ensemble can also be
observed on its reliability index of $0.3217$ and coverage value of
$61.21\,\%$, where the latter should be compared to the nominal
coverage of $83.33\,\%$ ($10/12$).

\section{Ensemble Model Output Statistics}
  \label{sec:sec3}
As mentioned in the Introduction, the EMOS predictive PDF
of a univariate weather quantity is a single parametric density
function, where the parameters depend on the ensemble members. In case
of temperature and pressure the normal distribution is a reasonable
choice \citep{grwg}, while for non-negative variables such as wind
speed, a skewed distribution is required. A popular candidate is the
Weibull distribution \citep[see, e.g.,][]{jhmg}, gamma or
log-normal distributions are also in use \citep{gtpf}, while \citet{tg}
suggested an EMOS model based on truncated normal distribution with a
cut-off at zero.

Let \ $f_1,f_2,\ldots ,f_M$ \ denote the ensemble of
distinguishable forecasts of wind speed for a given location
and time. This means that each ensemble member can be identified and
tracked, which holds for example for the UWME (see Section
\ref{subs:subs2.1}).

However, most of the currently used ensemble
prediction systems incorporate ensembles where at least some
members are statistically indistinguishable. 
Such ensemble systems are
usually producing initial conditions based on algorithms, which are
able to find the fastest growing perturbations indicating the directions
of the largest uncertainties. In most cases these initial
perturbations are  further enriched by perturbations simulating model
uncertainties as well. Examples in the paper at hand are the ECMWF 
ensemble and the ALADIN-HUNEPS ensemble described in Sections
\ref{subs:subs2.2} and \ref{subs:subs2.3}, respectively. In such cases one
usually has a control member (the one without any perturbation) and
the remaining ensemble members forming one or two exchangeable
groups.  

In what follows, if we have \ $M$ \ ensemble members divided
into \ $m$ \ exchangeable 
groups, where the \ $k$th \ group contains \ $M_k\geq 1$ \ ensemble
members ($\sum_{k=1}^mM_k=M$), \ notation \ $f_{k,\ell}$
\ is used for the  $\ell$th member of the $k$th group.

\subsection{Truncated normal model}
  \label{subs:subs3.1}

Denote by \ ${\mathcal N}^{\,0}\big(\mu,\sigma^2\big)$ \ the TN
distribution with location \ $\mu$, \ scale \ $\sigma>0$, \ and 
cut-off at zero having PDF 
\begin{equation*}
g(x\vert\, \mu,\sigma):=\frac{\frac
  1{\sigma}\varphi\big((x-\mu)/\sigma\big)}{\Phi\big(\mu/\sigma\big)
}, \quad x\geq 0, \qquad \text{and} \qquad g(x\vert\, \mu,\sigma):=0,
\quad \text{otherwise,}
\end{equation*}
where \ $\varphi$ \ and \ $\Phi$ \ are the PDF and the cumulative
distribution function (CDF) of the
standard normal distribution, respectively. 
The EMOS predictive distribution of wind speed \ $X$ \ proposed by
\citet{tg} is 
\begin{equation}
   \label{eq:eq3.1}
  {\mathcal N}_0\big(a_0+a_1f_1+ \cdots +a_Mf_M,b_0+b_1S^2\big) \qquad
  \text{with} \qquad S^2:=\frac 1{M-1}\sum_{k=1}^M\big (f_k-\overline f\big)^2,
\end{equation}
where \ $\overline f$ \ denotes the ensemble mean. 
Location parameters \ $a_0\in{\mathbb R}, \ a_1, \ldots, a_M\geq 0$ \ and
scale parameters \ $b_0,b_1\geq 0$ \ of model \eqref{eq:eq3.1} can be
estimated from the training data consisting of ensemble members and verifying
observations from the preceding \ $n$ \ days, by optimizing an
appropriate verification score (see Section \ref{subs:subs3.5}). 

If the ensemble can be divided into groups of
exchangeable members, ensemble members within a given group will get the
same coefficient of the location parameter \citep{frg} resulting in a
predictive distribution of the form 
\begin{equation}
   \label{eq:eq3.2}
  {\mathcal N}_0\bigg(a_0+a_1\sum_{\ell_1=1}^{M_1}f_{1,\ell_1}+ \cdots
  +a_m\sum_{\ell_m=1}^{M_m} f_{m,\ell_m},b_0+b_1S^2\bigg), 
\end{equation}
where again, \ $S^2$ \ denotes the ensemble variance. One might think
of taking into account the grouping also in modelling the variance of the
predictive PDF and use, e.g., the variance of the group means instead of
the ensemble variance \ $S^2$. \ However, practical tests show that
this (smaller) variance results in reduction of the predictive skill of
the model. 

\subsection{Log-normal model}
  \label{subs:subs3.2}

As an alternative to the TN model of Section \ref{subs:subs3.1} we
propose an EMOS approach based on an LN distribution. This
distribution has a heavier upper tail, and in this way it is more
appropriate to model high wind speed values. The PDF of the LN
distribution \ $\mathcal{LN}\big(\mu,\sigma\big)$ 
\ with location \ $\mu$ \ and shape \ $\sigma>0$ \ is
\begin{equation}
  \label{eq:eq3.3}
h(x\vert\, \mu,\sigma):=\frac
  1{x\sigma}\varphi\big((\log x-\mu)/\sigma\big), \quad x\geq 0,
  \qquad \text{and} \qquad h(x\vert\, \mu,\sigma):=0, 
\quad \text{otherwise,}
\end{equation}
while the mean \ $m$ \ and variance \ $v$ \ of this distribution are
\begin{equation*}
m={\mathrm e}^{\,\mu +\sigma^2/2}  \qquad \text{and} \qquad v={\mathrm
  e}^{2\mu +\sigma^2} \big ({\mathrm e}^{\sigma^2}-1\big), \qquad
\text{respectively.}  
\end{equation*} 
Further, since
\begin{equation}
  \label{eq:eq3.4}
\mu =\log \bigg(\frac {m^2}{\sqrt{v+m^2}}\bigg) \qquad \text{and}
\qquad \sigma=\sqrt{\log \Big (1+ \frac v{m^2}\Big)},
\end{equation}
an LN distribution can also be parametrized by these quantities.
In our EMOS approach \ $m$ \ and \ $v$ \ are affine functions of the
ensemble members and ensemble variance, respectively, that is
\begin{equation}
\label{eq:eq3.5}
m=\alpha_0+\alpha_1f_1+ \cdots +\alpha_Mf_M \qquad\text{and}\qquad
v=\beta_0+\beta_1 S^2.
\end{equation}
Similar to the TN model, to obtain the values of mean and variance
parameters \ $\alpha_0\in{\mathbb R}, \ \alpha_1, \ldots, \alpha_M\geq 0$ \ and
\ $\beta_0,\beta_1\geq 0$, \ respectively, one has to perform an
optimum score estimation based on some verification
measure. Obviously, for the case of exchangeable ensemble members
instead of \eqref{eq:eq3.5} we have
\begin{equation}
  \label{eq:eq3.6}
m=\alpha_0+\alpha_1\sum_{\ell_1=1}^{M_1}f_{1,\ell_1}+ \cdots
  +\alpha_m\sum_{\ell_m=1}^{M_m} f_{m,\ell_m} \qquad\text{and}\qquad
  v=\beta_0+\beta_1 S^2.
\end{equation}

\subsection{Combined model}
  \label{subs:subs3.3}

To combine the advantageous properties of TN and LN approaches,
following \citet{lt}, we also investigate a regime-switching
method. Depending on the value of the ensemble median \ $f_{med}$ \
we consider either a TN or an LN based EMOS model. Given a threshold
\ $\theta>0$, \ the EMOS predictive distribution is \ ${\mathcal
  N}^{\,0}\big(\mu_{TN},\sigma_{TN}^2\big)$ \ if \ $f_{med}<\theta$ \
and \ $\mathcal{LN}\big(\mu_{LN},\sigma_{LN}\big)$, \
otherwise. Model parameters \ $\mu_{TN}$ \ and \ $\sigma_{TN}$ \
depend on the ensemble forecast according to \eqref{eq:eq3.1} or
\eqref{eq:eq3.2}, while the expressions for \ $\mu_{LN}$ \ and \
$\sigma_{LN}$ \ can be obtained form \eqref{eq:eq3.5} or
\eqref{eq:eq3.6} via transformation \eqref{eq:eq3.4}. For training the
combined model we propose two different methods. If the training data
set is large enough, that is many forecast cases belong to each day to
be investigated, the LN model is trained using only ensemble forecasts where \
$f_{med}\geq\theta$, \  while forecasts with ensemble median under the
threshold are used to train the TN model. This technique is applied
for calibrating the UWME and the ECMWF ensemble forecasts, see Sections
\ref{subs:subs4.1} and \ref{subs:subs4.2}, respectively. However, e.g.,
in case of the ALADIN-HUNEPS ensemble one has only 10 observation
stations, so there are not enough data for separate training of the
component models. In such
situations one might utilize the same training data set
both for the TN and for the LN predictive distribution and then
choose between these two models according to the value of the ensemble
median. This particular idea is applied in Section \ref{subs:subs4.3} for the
ALADIN-HUNEPS forecasts.

\subsection{General Extreme Value model}
  \label{subs:subs3.4}

In Section \ref{sec:sec4} the predictive performances of the LN and
TN-LN mixture models are compared to those of the GEV and TN-GEV
mixture models of \citet{lt}. The CDF of a GEV distribution \
$\mathcal{GEV}\big(\mu,\sigma,\xi\big)$ \ with location \ $\mu $, \
scale \ $\sigma>0$ \ and shape \ $\xi$ \ equals
\begin{equation*}
G(x\vert\, \mu,\sigma,\xi ):=\begin{cases}
\exp\Big(-\big[1+\xi(\frac{x-\mu}{\sigma})\big]^{-1/\xi}\Big), & \
\xi\ne 0; \\
\exp\Big(-\exp\big(-\frac{x-\mu}{\sigma}\big)\Big), & \
\xi= 0,
\end{cases} 
\end{equation*}  
if \ $1+\xi(x-\mu)/\sigma> 0$, \ and zero otherwise. This definition
shows the main disadvantage of using a GEV distribution for modelling
wind speed, namely, there is a positive probability for a GEV
distributed random variable to be negative. 

For calibrating ECMWF ensemble forecasts of wind speed over Germany
\citet{lt} suggest to model location and scale parameters by
\begin{equation}
  \label{eq:eq3.7}
\mu=\gamma_0+\gamma_1f_1+ \cdots +\gamma_Kf_K \qquad \text{and} \qquad
\sigma=\sigma_0+\sigma_1\overline f,
\end{equation}
while the shape parameter \ $\xi$ \ is considered to be independent of the
ensemble. In general, one can also incorporate the ensemble variance 
 into the models of location and scale . However, preliminary studies
showed that model \eqref{eq:eq3.7} is also a reasonable choice for the
UWME and the ALADIN-HUNEPS ensemble. Further, the components of the
TN-GEV mixture model for the various 
ensemble forecasts are trained as described in Section \ref{eq:eq3.2}.

\subsection{Parameter estimation}
  \label{subs:subs3.5}

The aim of probabilistic forecasting is to obtain calibrated and sharp
predictive distributions of future weather quantities
\citep{gbr}. This goal should also be addressed in the  
choice of the scoring rule to be optimized in order to obtain the estimates
of parameters of different EMOS models. For evaluating density
forecasts the most popular scoring rules are the logarithmic score
\citep{grjasa}, i.e. the negative logarithm of the predictive PDF
evaluated at the verifying observation, and the continuous ranked 
probability score \citep[CRPS;][]{grjasa,wilks}. Given a predictive CDF
\ $F(y)$ \ and an observation \ $x$, \ the CRPS is defined as 
\begin{equation}
  \label{eq:eq3.8}
\crps\big(F,x\big):=\int_{-\infty}^{\infty}\big (F(y)-{\mathbbm 
  1}_{\{y \geq x\}}\big )^2{\mathrm d}y={\mathsf E}|X-x|-\frac 12
{\mathsf E}|X-X'|, 
\end{equation}
where \ ${\mathbbm 1}_H$ \ denotes the indicator of a set \ $H$, \
while \ $X$ \ and \ $X'$ \ are independent random variables with CDF \
$F$ \ and finite first moment. We remark that the CRPS can be expressed in
the same unit as the observation. Both the logarithmic score and the CRPS
are proper scoring rules which are 
negatively oriented, that is, the smaller the better.  In this way the
optimization with respect to the logarithmic score gives back the
maximum likelihood (ML) estimates of the parameters.

Short calculation shows that the CRPS corresponding to the CDF \
$\mathcal G$ \ of a TN 
distribution \  ${\mathcal N}^{\,0}\big(\mu,\sigma^2\big)$ \ can be
given in a closed form \citep[see, e.g.,][]{tg}, namely 
\begin{align*}
\crps\big(\mathcal G,x\big)=\sigma \bigg[\Phi\big(\mu/\sigma\big)
\bigg]^{-2}&\, 
\bigg[\frac{x-\mu}{\sigma} \Phi\big(\mu/\sigma\big) \Big (2
\Phi\big((x-\mu)/\sigma\big)+\Phi\big(\mu/\sigma\big)-2\Big) \\
&+2\varphi\big((y-\mu)/\sigma\big)\Phi\big(\mu/\sigma\big) 
-\frac 1{\sqrt{\pi}}\Phi\big(\sqrt{2}\mu/\sigma\big)\bigg]. 
\end{align*}

In case of the LN model one faces a similar situation, straightforward
calculations verify
\begin{align*}
\crps\big(\mathcal H,x\big)\!=\!x \Big[2\Phi\big((\log
x\!-\!\mu)/\sigma\big)\!-\! 1\Big] \!\!-\! 2{\mathrm e}^{\mu+\sigma
  ^2/2}\Big[\Phi\big((\log 
x\!-\!\mu)/\sigma\!-\!\sigma \big)\!+\!\Phi \big(\sigma/\sqrt
2\big)\!-\!1\Big], 
\end{align*}
where \ $x\geq 0$ \ and \ $\mathcal H$ \ is the CDF corresponding to the PDF
\eqref{eq:eq3.3} of \  $\mathcal{LN}\big(\mu,\sigma\big)$. \
Obviously, with the help of transformations \eqref{eq:eq3.4}, \ 
$\crps\big(\mathcal H,x\big)$ \ can also be expressed as a function of
the mean \ $m$ \ and variance \ $v$ \ of the LN distribution
$\mathcal{LN}\big(\mu,\sigma\big)$.  

Now, following the ideas of \citet{grwg} and \citet{tg}, both for the
TN and the LN model we estimate model parameters by minimizing the
mean CRPS of the predictive distributions and validating observations
corresponding to the forecast cases of the training period, while for
the GEV model the ML method suggested by \citet{lt} is applied.

\section{Results}
 \label{sec:sec4}
As mentioned in the Introduction, the predictive performances of the LN
model and of the TN-LN combined model (see Sections \ref{subs:subs3.2}
and \ref{subs:subs3.3}, respectively) are tested on the eight-member
UWME, on the fifty-member ECMWF ensemble and on the ALADIN-HUNEPS
ensemble of the HMS. The obtained results are compared to the fits of the TN,
GEV and TN-GEV combined models investigated by \citet{lt}, and to the
verification scores of the raw ensemble. We also consider the scores
corresponding to climatological forecasts which can be defined as
forecasts calculated from observations in the training period
used as an ensemble.

The goodness of fit of a calibrated forecast in terms of probability
distributions is quantified with the help of the mean CRPS defined in
Section \ref{subs:subs3.4}. For the raw ensemble and climatology in
\eqref{eq:eq3.8} 
the empirical CDF replaces the EMOS predictive CDF. In order to evaluate
forecasts for high wind speeds we also consider the threshold-weighted 
continuous ranked probability score 
(twCRPS) 
\begin{equation*}
\twcrps\big(F,x\big):=\int_{-\infty}^{\infty}\big (F(y)-{\mathbbm 
  1}_{\{y \geq x\}}\big )^2\omega(y){\mathrm d}y
\end{equation*}
introduced by \citet{gr}, where \ $\omega(y)\geq 0$ \ is a weight
function. Obviously, case \ $\omega(y)\equiv 1$ \ corresponds to the
traditional CRPS defined by  \eqref{eq:eq3.8}, while to address wind
speeds above a given threshold \ $r$ \ one may set \ $\omega
(y)={\mathbbm 1}_{\{y \geq r\}}$.  \ In our study we consider
threshold values approximately corresponding to the 90th, 95th and
99th percentiles of the wind speed observations. Further, in order to
quantify the improvement in twCRPS with respect to a reference predictive
CDF \ $F_{ref}$ \ we make use of the threshold-weighted continuous
ranked probability skill score (twCRPSS) defined as \citep[see,
e.g.,][]{lt} 
\begin{equation*}
\twcrpss\big(F,x\big):=1-\frac{\twcrps\big(F,x\big)}{\twcrps\big(F_{ref},x\big)}.
\end{equation*}
This score is obviously positively oriented, and as a reference we
always use the predictive CDF corresponding to the TN model.

Finally, for each EMOS model we investigate the
coverage and average width of the central prediction interval
corresponding to the nominal coverage of 
the raw ensemble (UWME: $77.78\,\%$; ECMWF: $96.08\,\%$; ALADIN-HUNEPS:
$83.33\,\%$), where the coverage of a \ $(1-\alpha)100 \,\%, \ \alpha \in
(0,1),$ \ central prediction 
interval is the proportion of validating observations located between
the lower and upper \ $\alpha/2$ \ quantiles of the predictive
distribution. For a calibrated predictive PDF this value should be
around \ $(1-\alpha)100 \,\%$ \ and the proposed choices of \ $\alpha$
\ allow direct comparisons to the raw ensembles. 

A continuous counterpart of the verification rank histogram (see
Figure \ref{fig:fig1}) of the raw ensemble is the probability integral
transform (PIT) histogram of the predictive distribution. The PIT is 
the value of the predictive CDF evaluated at
the verifying observation \citep{rgbp}, and the PIT histogram provides a good
measure about the possible improvements of the under-dispersive
character of the raw ensemble.  The closer the histogram is to the
uniform distribution, the better is the calibration. 

\begin{figure}[t]
\subfigure[$\hspace{-1.25cm}$]{\includegraphics[width=0.33
  \textwidth]{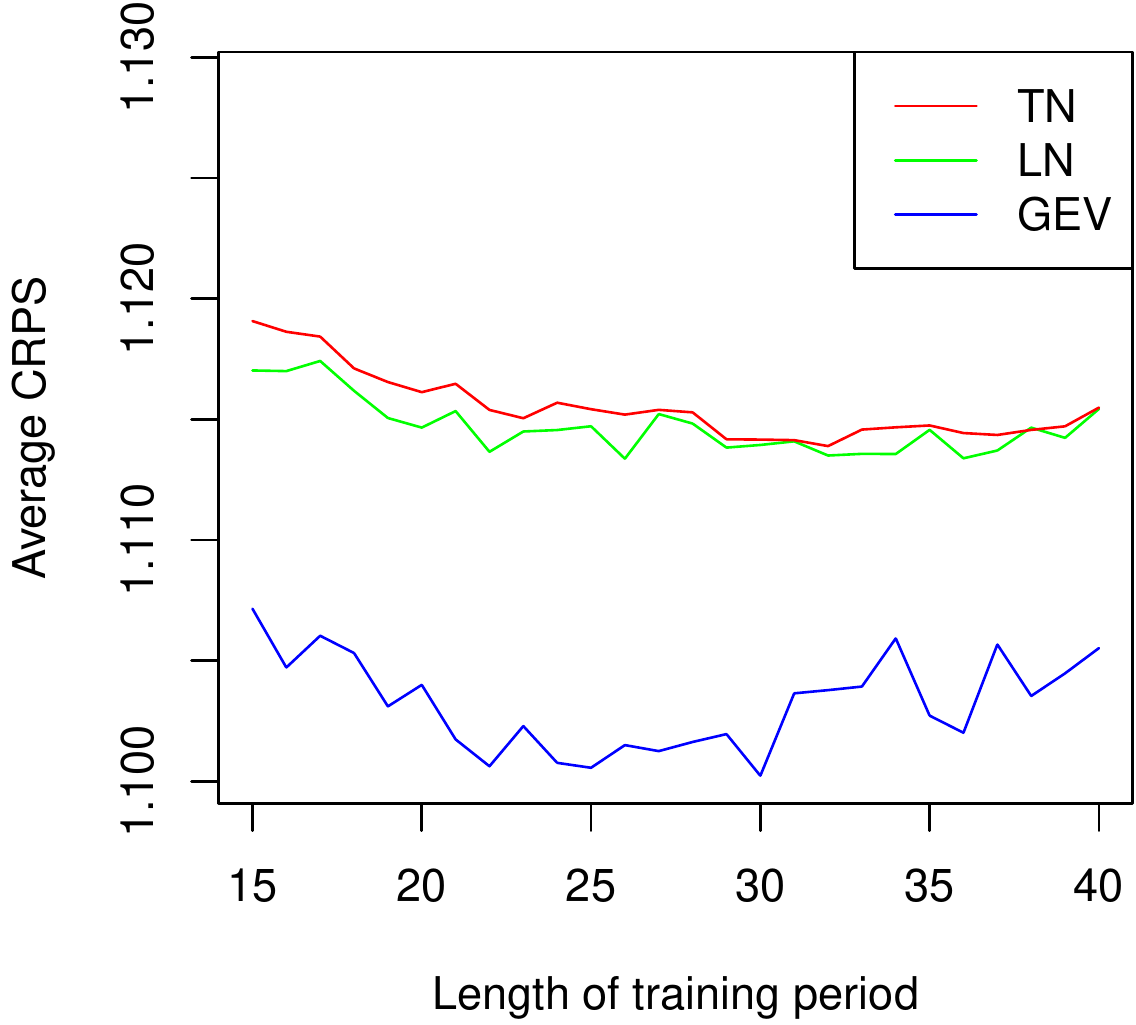}}\hfill  
\subfigure[$\hspace{-1.25cm}$]{\includegraphics[width=0.33
  \textwidth]{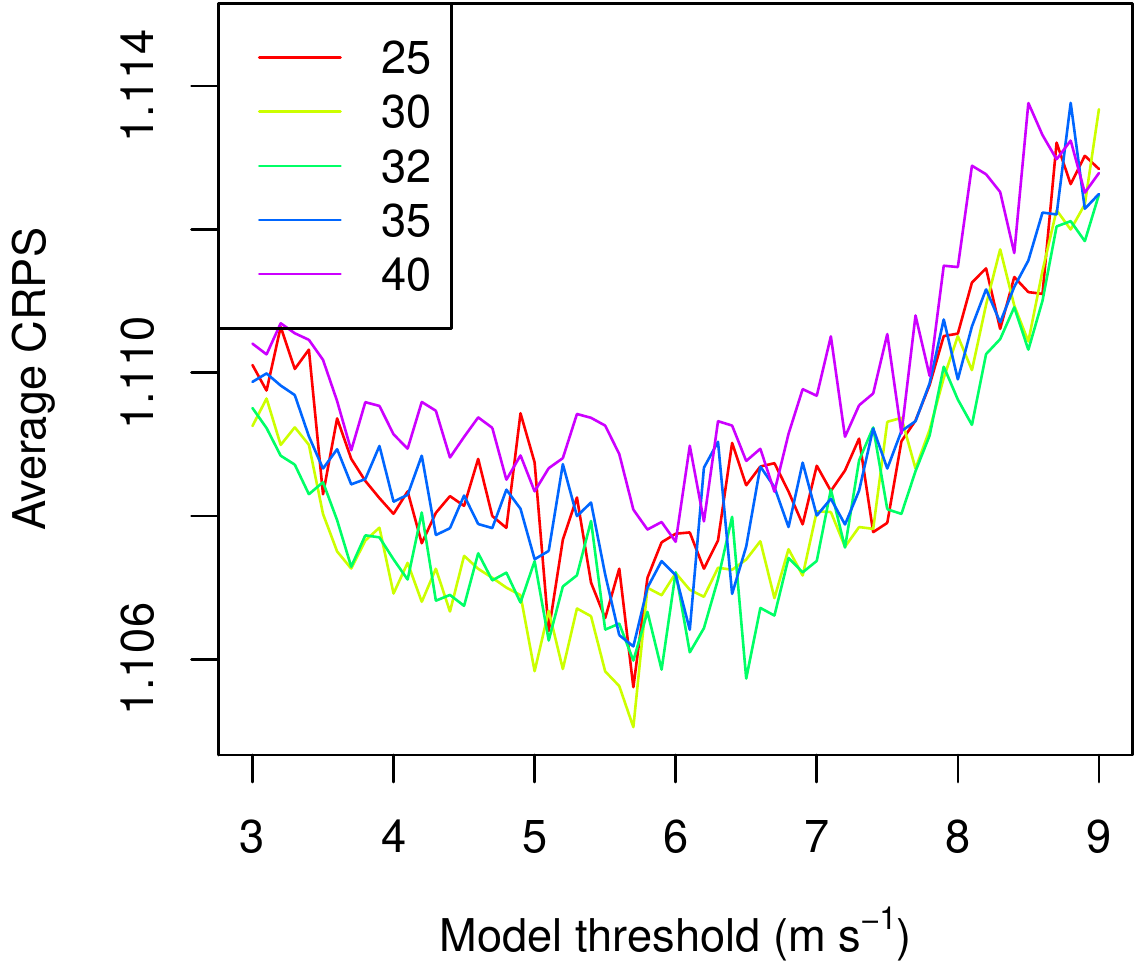}}\hfill 
\subfigure[$\hspace{-1.25cm}$]{\includegraphics[width=0.33
  \textwidth]{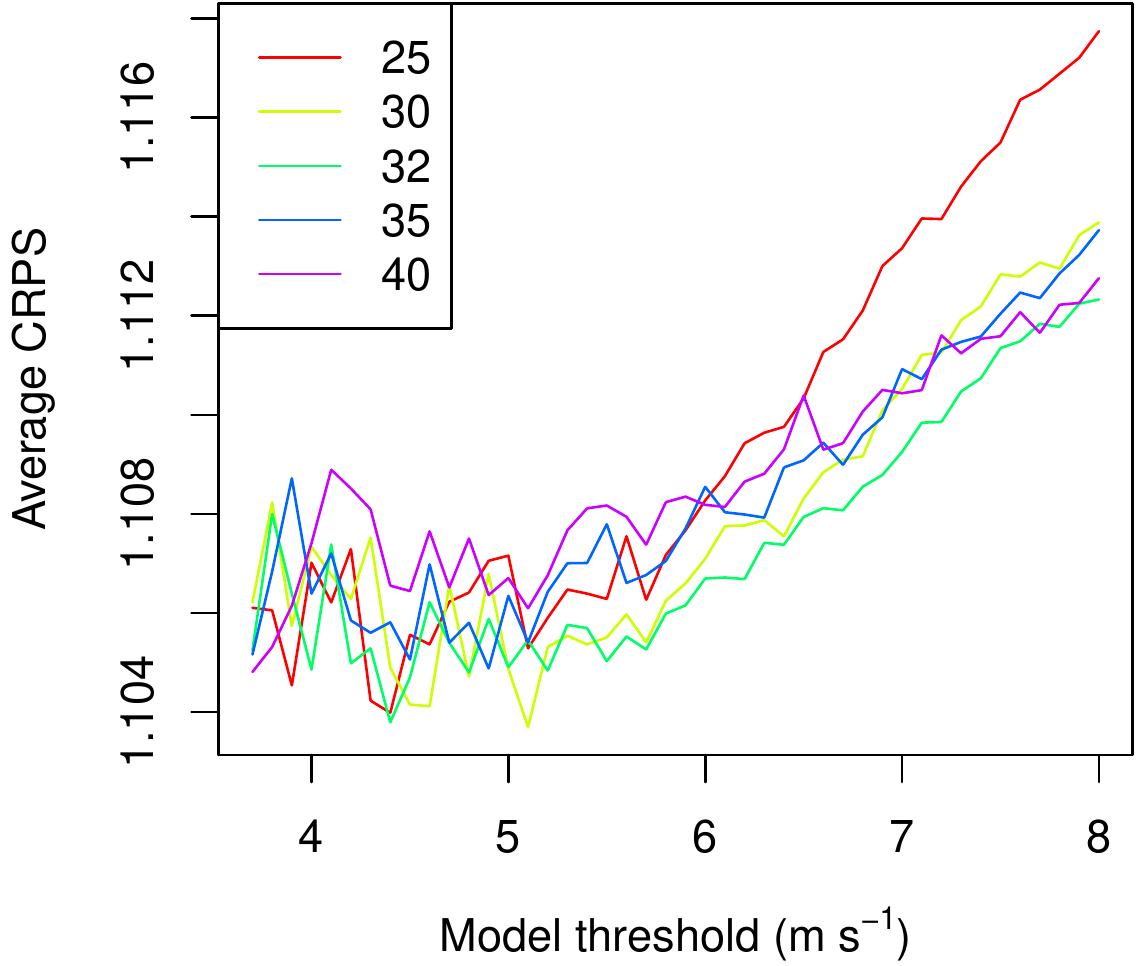}} 
\caption{Mean CRPS values of the (a) EMOS predictive distributions for various 
  training period lengths; (b) TN-LN mixture models corresponding to different
  training period lengths as functions of the threshold; (c) TN-GEV
  mixture models corresponding to different 
  training period lengths as functions of the threshold for the UWME.}
\label{fig:fig2}
\end{figure}

As point forecasts we consider EMOS and ensemble medians and means,
which are evaluated with the use of mean absolute errors (MAEs) and
root mean squared errors (RMSEs). Note that MAE is optimal for the
median, while RMSE is optimal for the mean forecasts
\citep{gneiting11,pinhag}.

\subsection{University of Washington Mesoscale Ensemble}
  \label{subs:subs4.1}

\begin{figure}[t!]
\begin{center}
\includegraphics[width=\textwidth]{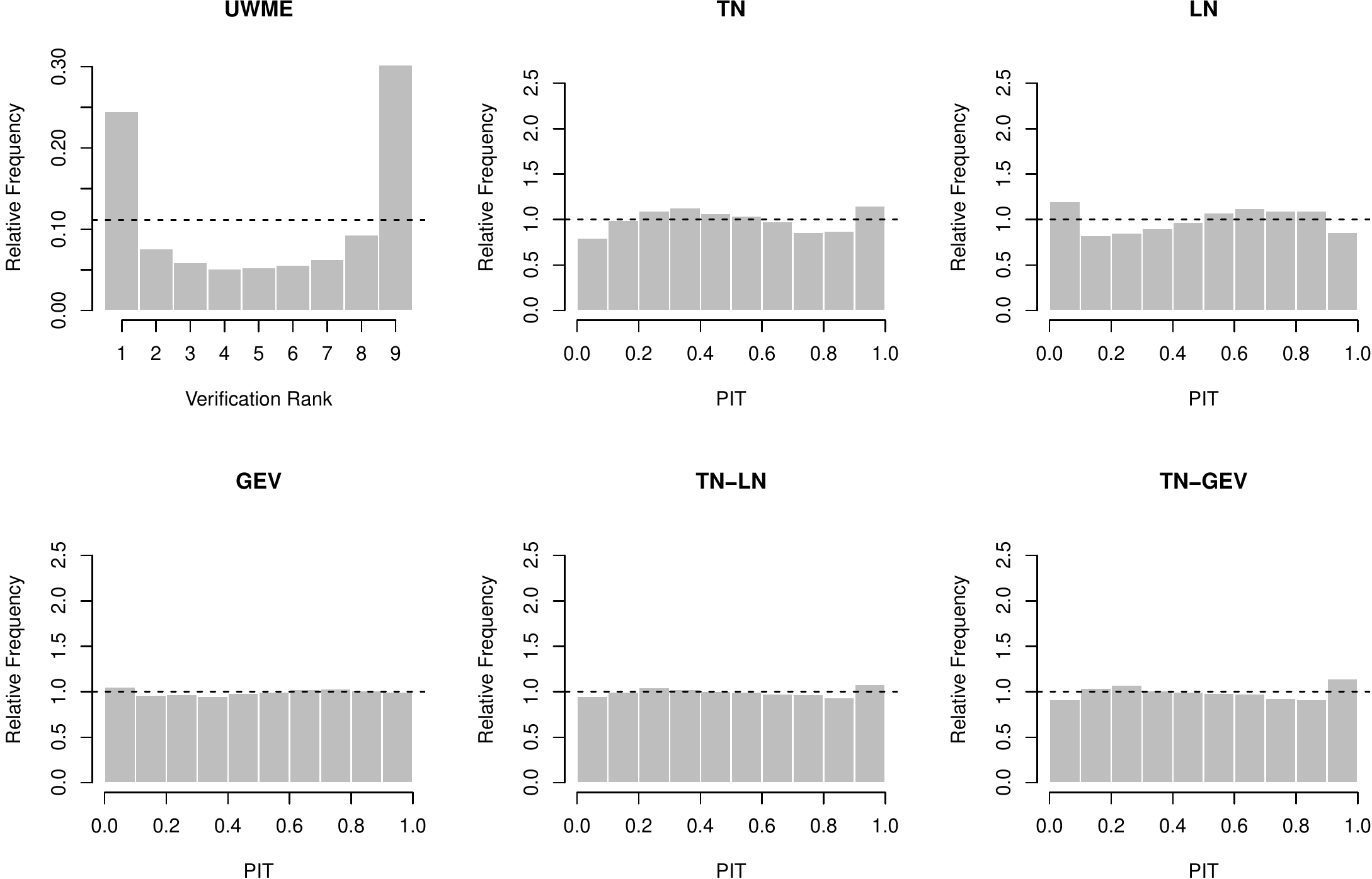}
\caption{Verification rank histogram of the raw ensemble and PIT
  histograms of the EMOS post-processed forecasts for the UWME.}   
\label{fig:fig3}
\end{center}
\end{figure} 

As the eight members of the UWME are non-exchangeable, the dependencies
of location and scale parameters of the TN and GEV models on the
ensemble members are specified by \eqref{eq:eq3.1}, \eqref{eq:eq3.7},
respectively, while the mean and variance of the LN model are linked
to the ensemble according to \eqref{eq:eq3.5}.

As a first step we determine the optimal length of the rolling
training period valid
for all models and the optimal threshold values for TN-LN and TN-GEV
mixtures. Figure \ref{fig:fig2}a shows the mean CRPS values of all
three models as functions of the training period length varying from 15
to 40 days. The mean CRPS of the GEV model takes its minimum at day 30
and this training period length seems reasonable for the other two
models, too. This particular length of the training period is also supported by
Figure \ref{fig:fig2}b showing the mean CRPS values of the TN-LN
mixture model as
function of the threshold \ $\theta$ \ for various training period
lengths. For this mixture model the optimal threshold is $5.7$ m/s,
while for the TN-GEV models similar arguments lead us to a threshold
of $5.2$ m/s, see Figure \ref{fig:fig2}c. Using these parameter values
ensemble forecasts for the calendar year 2008 are calibrated. In case
of the two regime-switching models an LN distribution is used in
around one third, while a GEV distribution is applied
in about 40\,\% of the 27\,481 individual forecast cases. 

\begin{table}[t!]
\begin{center}
\begin{tabular}{|ll|c|c|c|c|c|c|c|c|} \hline
Forecast&&CRPS&\multicolumn{3}{c|}{
  twCRPS $(m/s)$}&MAE&RMSE&Cover.&Av. w.\\\cline{4-6}
&&$(m/s)$&$r\!=\!9$&$r\!=\!10.5$&$r\!=\!14$&
$(m/s)$&$(m/s)$&$(\%)$&$(m/s)$\\ \hline
TN&&1.114&0.150&0.074&0.010&1.550&2.048&78.65&4.67 \\
LN&&1.114&0.149&0.073&0.010&1.554&2.052&77.29&4.69 \\
TN-LN,&$\!\!\!\!\theta\!=\!5.7$&1.105&0.149&0.073&0.010&1.550&2.050&77.73&
4.64 \\
GEV&&1.100&0.145&0.072&0.010&1.554&2.047&77.20&4.69 \\
TN-GEV,&$\!\!\!\!\theta\!=\!5.2$&1.105&0.145&0.072&0.010&1.555&2.055&77.20&
4.60 \\\hline
Ensemble&&1.353&0.175&0.085&0.011&1.655&2.169&45.24&2.53 \\
Climatology&&1.412&0.173&0.081&0.010&1.987&2.629&81.10&5.90 \\\hline
\end{tabular} 
\caption{Mean CRPS, mean twCRPS for various thresholds \ $r$, \ MAE of
  median and RMSE of mean forecasts and coverage and 
  average width of $77.78\,\%$ central prediction intervals for the
  UWME.} \label{tab:tab1} 
\end{center}
\end{table}

\begin{figure}[b!]
\begin{center}
\includegraphics[width=0.45\textwidth]{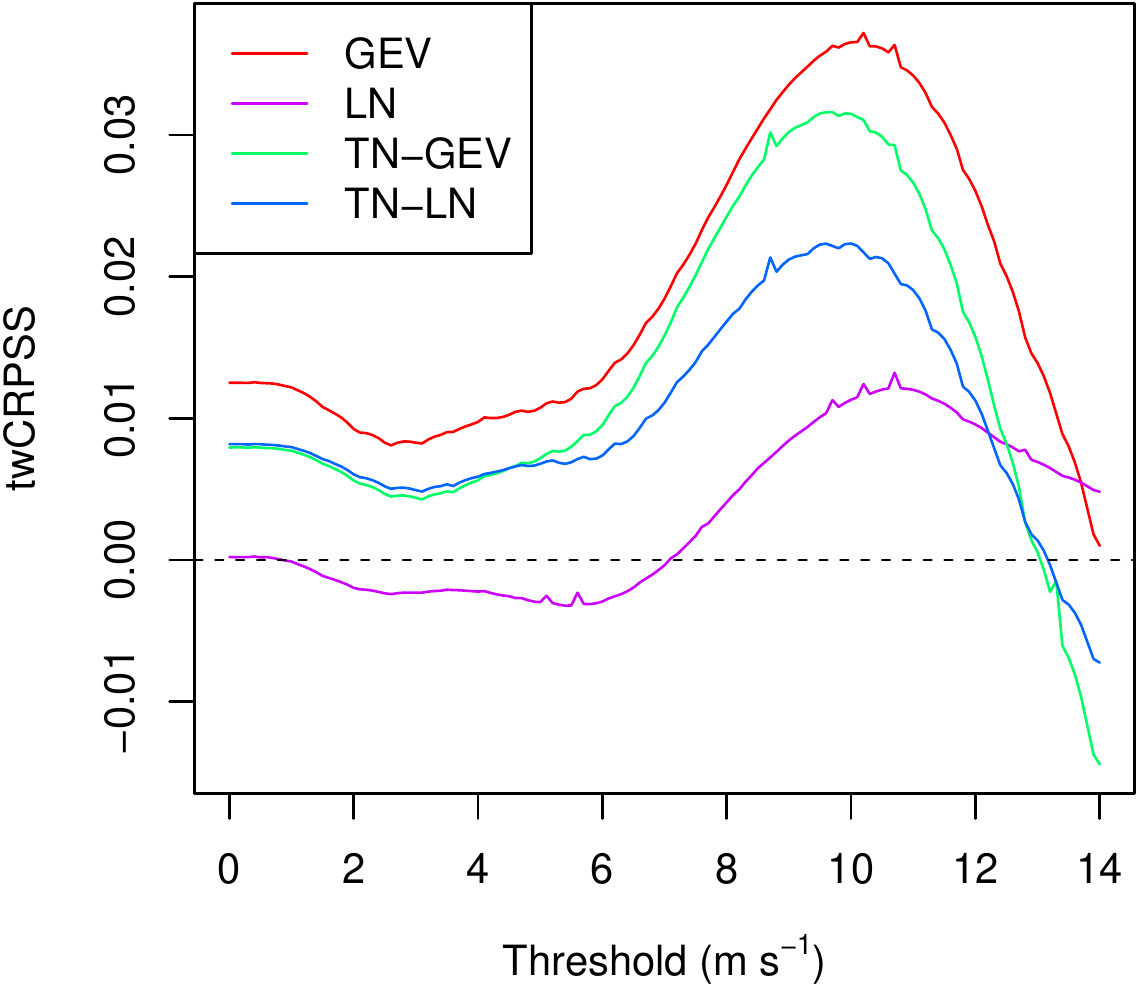}
\caption{twCRPSS values for the UWME with TN as reference model.}   
\label{fig:fig4}
\end{center}
\end{figure}

Consider first the PIT histograms of the investigated
EMOS models that are displayed in Figure \ref{fig:fig3}. A comparison
to the verification rank histogram of the raw ensemble shows that
post-processing significantly improves the statistical calibration of
the forecasts. However, one should admit that, e.g., the
Kolmogorov-Smirnov test rejects the uniformity of PIT values in all
cases, the highest $p$-value, corresponding to the GEV model, is
$0.0049$.

In Table \ref{tab:tab1} scores for different probabilistic forecasts
are given together with the average width and coverage of $77.75\,\%$ 
central prediction intervals. Verification measures of probabilistic
forecasts and point forecasts calculated using TN, LN and GEV models
and TN-LN  ($\theta=5.7$ m/s) and TN-GEV  ($\theta=5.2$ m/s) mixture
models are compared to the corresponding 
measures calculated for the raw ensemble and climatological
forecasts.  By examining these results, one can clearly observe 
the obvious advantage of  
post-processing with respect to the raw ensemble or to the
climatology. This is quantified 
in decrease of CRPS, MAE and RMSE values and in a significant improvement
in the coverage of the 
$77.78\,\%$ central prediction intervals. On the other hand, the post-processed
forecasts are less sharp than the ones calculated from the raw
ensemble, however, this fact is coming from the small
dispersion of the UWME, as 
also seen in the verification rank histogram of Figure
\ref{fig:fig1}a. 

From the five competing models the GEV method produces the smallest
CRPS and RMSE 
values and the lowest twCRPS scores for all three thresholds reported,
while the best coverage and MAE value correspond to the TN-LN 
mixture model. However, the superiority of the GEV model is not
surprising after examining Figure \ref{fig:fig4} showing the twCRPSS
values of GEV, LN, TN-GEV and TN-LN EMOS methods with respect to the
reference TN model as functions of the threshold. As the GEV model
outperforms the TN model (and the other three, as well) at all investigated
thresholds, combining the two methods does not result in an increase
in the predictive skill. Hence, one can conclude that in case of the
UWME data the GEV model has the best overall performance, but one
should also remark that for this model the mean (maximal) probability
of forecasting a negative wind speed is around $0.05\,\% \ (4 \,\%)$.

\begin{figure}[t!]
\subfigure[$\hspace{-1.25cm}$]{\includegraphics[width=0.33
  \textwidth]{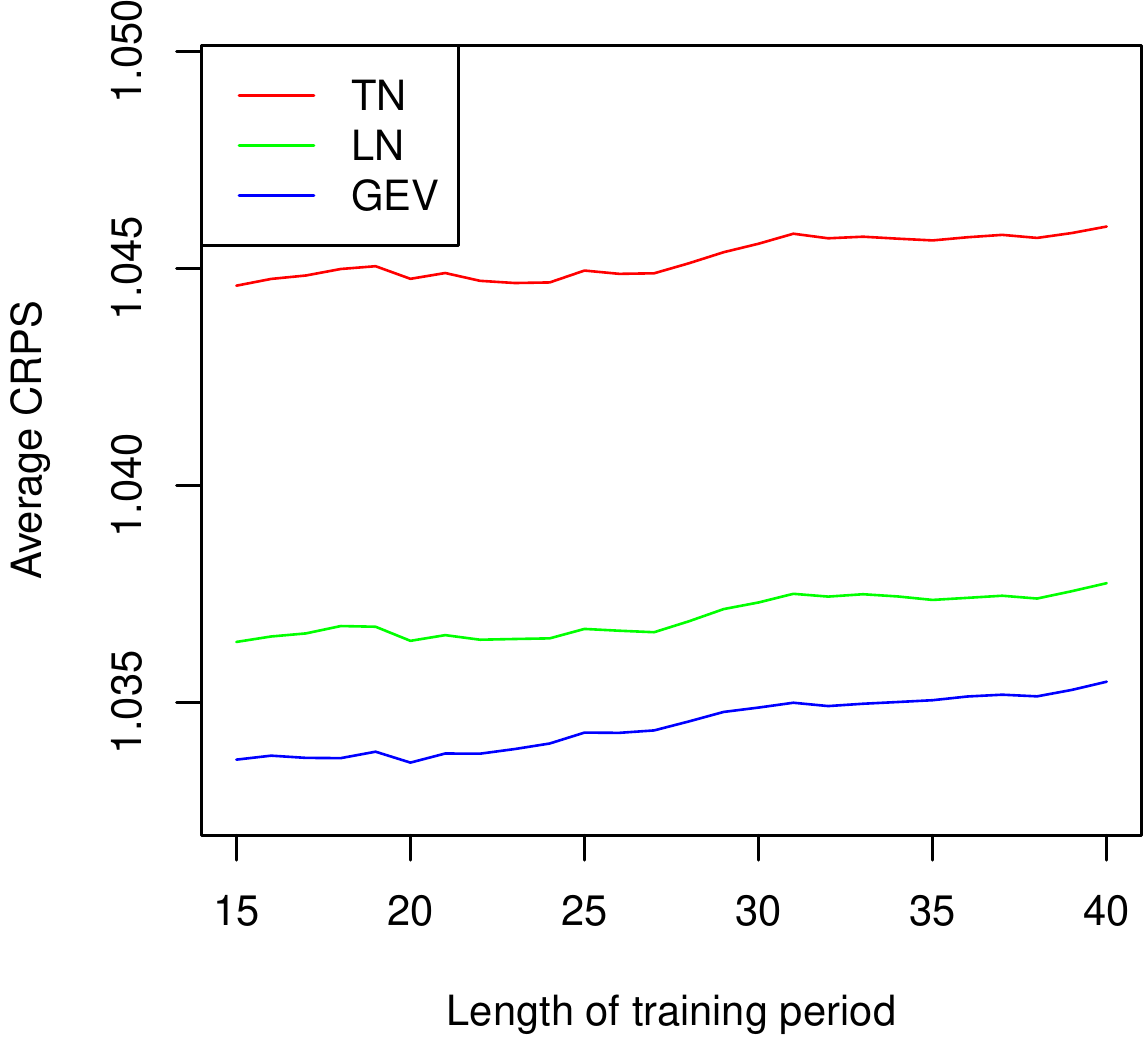}}\hfill  
\subfigure[$\hspace{-1.25cm}$]{\includegraphics[width=0.33
  \textwidth]{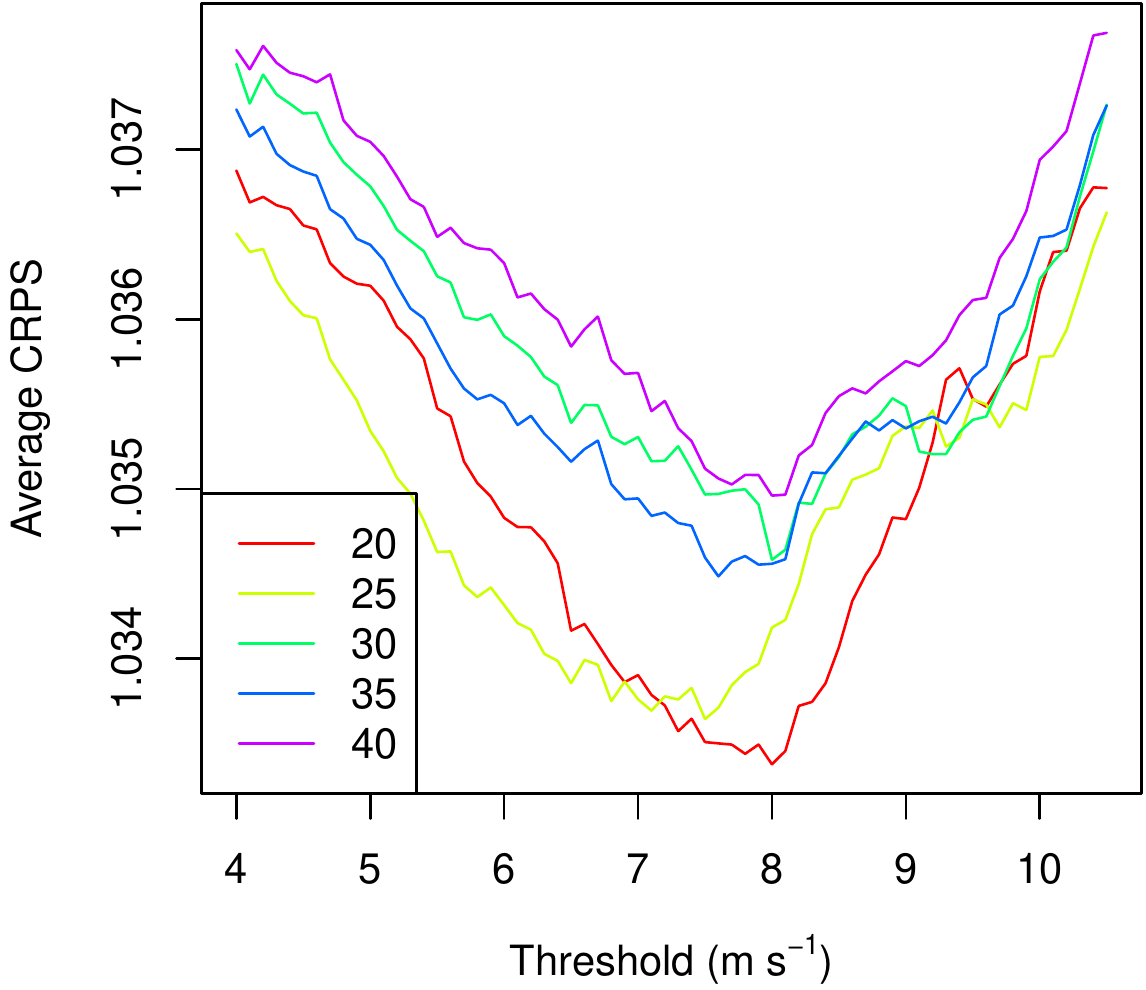}}\hfill 
\subfigure[$\hspace{-1.25cm}$]{\includegraphics[width=0.33
  \textwidth]{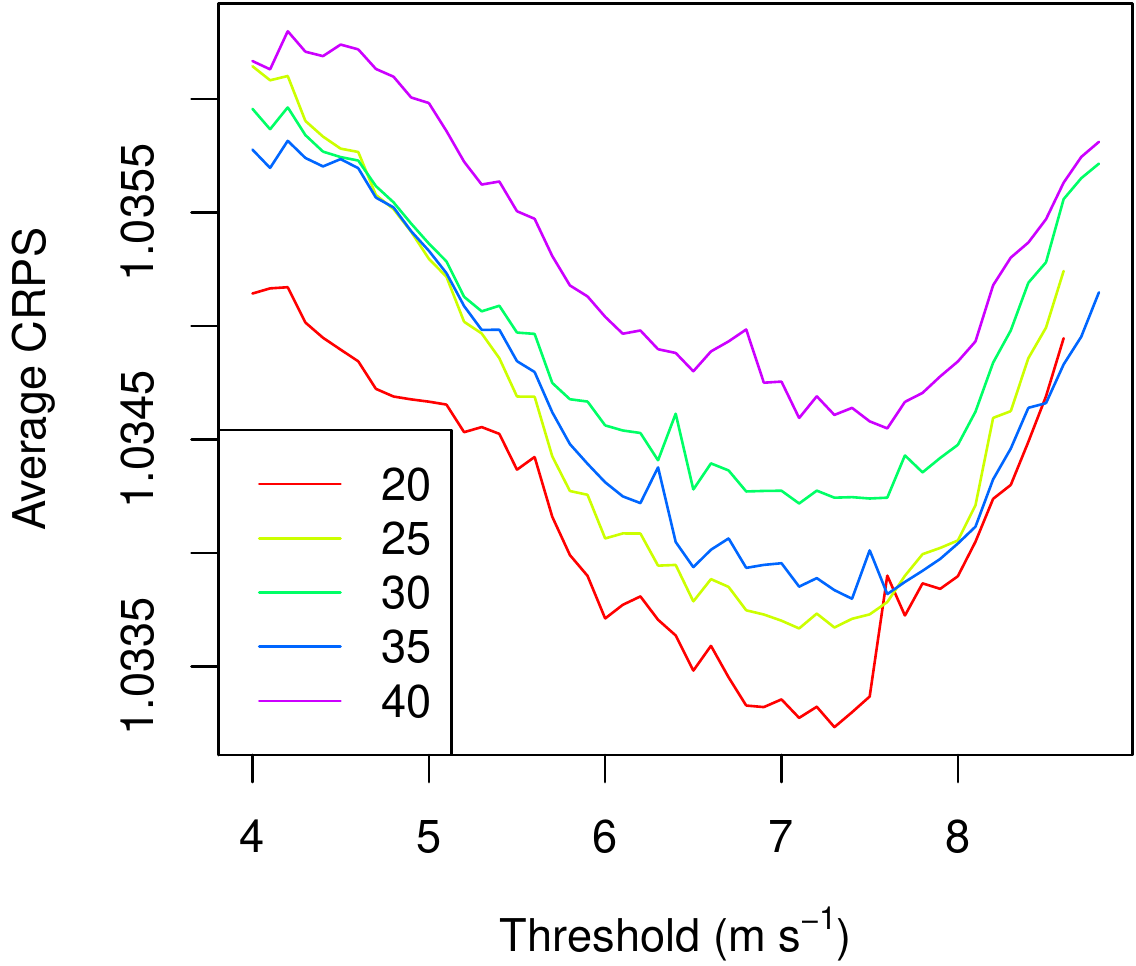}} 
\caption{Mean CRPS values of the (a) EMOS predictive distributions for various 
  training period lengths; (b) TN-LN mixture models corresponding to different
  training period lengths as functions of the threshold; (c) TN-GEV
  mixture models corresponding to different 
  training period lengths as functions of the threshold for the ECMWF
  ensemble.} 
\label{fig:fig5}
\end{figure}

\subsection{ECMWF ensemble}
  \label{subs:subs4.2}

The ECMWF ensemble consists of one group of 50 exchangeable
members. The parameters of the TN and the LN model are thus linked to
the ensemble according to \eqref{eq:eq3.2} and \eqref{eq:eq3.6} with
$m=1$. Following \citet{lt}, the location and scale parameter of the
GEV model are given as specified in \eqref{eq:eq3.7} with
$\gamma_1,\dots,\gamma_K$ restricted to be equal. 

To determine the optimal length of the training period for all models
and the optimal model thresholds for the combination models we proceed
as for the UWME ensemble and compute the average CRPS over a range of
lengths of training periods and choices for the model threshold
$\theta$, see Figure \ref{fig:fig5}. Figure
\ref{fig:fig5}a suggests a training period of 20 days
for all models, while Figures \ref{fig:fig5}b and
\ref{fig:fig5}c suggest a model threshold of $\theta
= 8.0$ m/s and $\theta = 7.3$ m/s for the TN-LN model and the TN-GEV
model, respectively. With these threshold values, an LN distribution
is used in around 14\,\% of the forecast cases in the verification set,
and a GEV distribution is used in around 19\,\% of the forecast cases.  
  
\begin{figure}[t]
\begin{center}
\includegraphics[width=\textwidth]{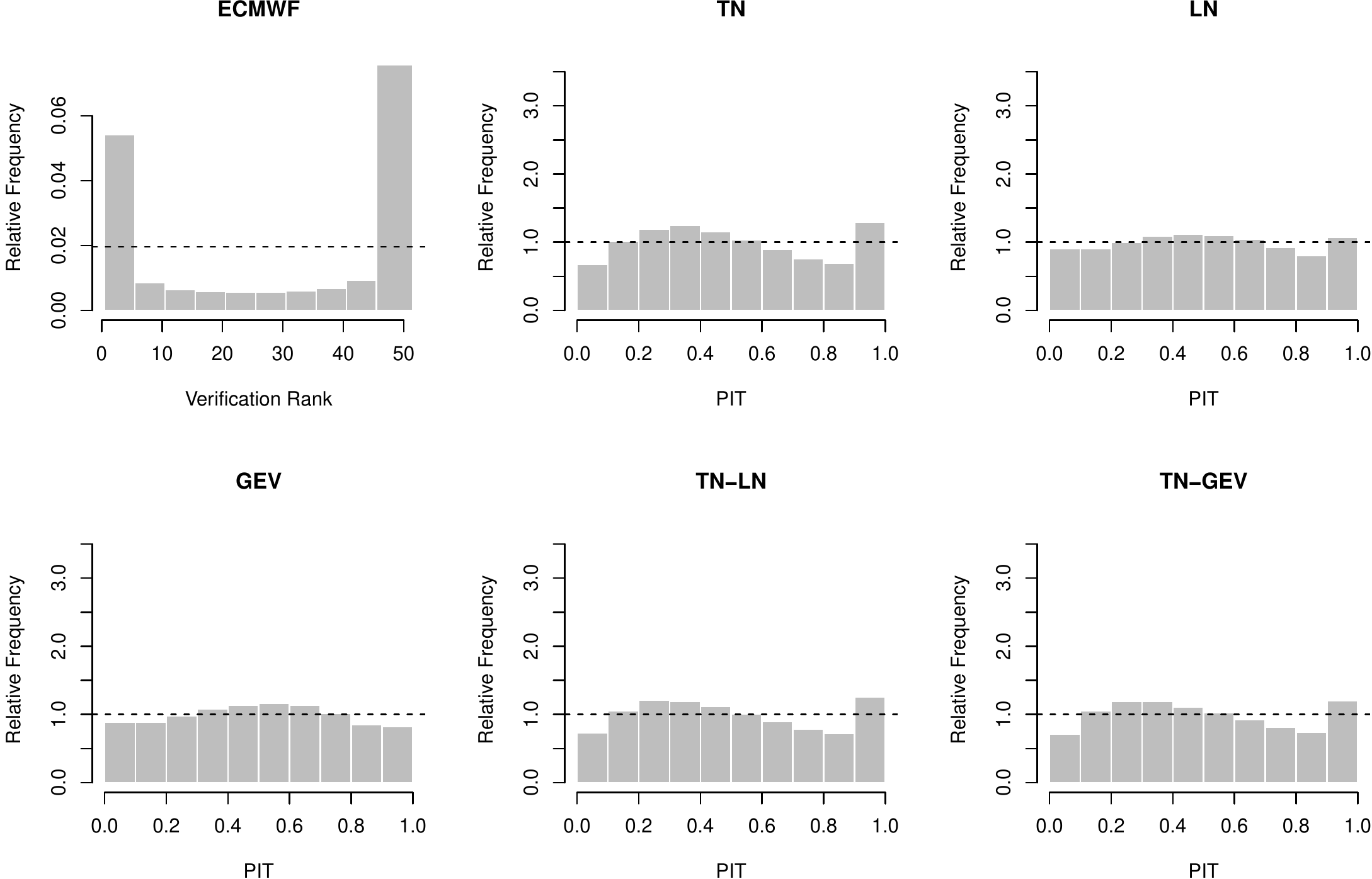}
\caption{Verification rank histogram of the raw ensemble and PIT
  histograms of the EMOS post-processed forecasts.}   
\label{fig:fig6}
\end{center}
\end{figure} 

Figure \ref{fig:fig6} showing the verification rank histogram of the
raw ensemble and the PIT histograms of the various predictive distributions
illustrates that all post-processing  
methods significantly increase the calibration of the ensemble. While
the tails of the TN model appear to be slightly too light, the PIT
histogram of the GEV model is gradually over-dispersive with minimally
too heavy tails. The smallest deviations from uniformity are obtained
for the LN model. The PIT histograms for the combination models
resemble the PIT histogram of the TN model with small improvements at
higher PIT values. Note that Kolmogorov-Smirnov tests reject the uniformity of
the PIT values for all five models.   

\begin{table}[t!]
\begin{center}
\begin{tabular}{|ll|c|c|c|c|c|c|c|c|} \hline
Forecast&&CRPS&\multicolumn{3}{c|}{
  twCRPS $(m/s)$}&MAE&RMSE&Cover.&Av.w.\\\cline{4-6}
&&$(m/s)$&$r\!=\!10$&$r\!=\!12$&$r\!=\!15$&
$(m/s)$&$(m/s)$&$(\%)$&$(m/s)$\\ \hline
TN&&1.045&0.200&0.110&0.042&1.388&2.148&92.19&6.39 \\
LN&&1.037&0.198&0.109&0.042&1.386&2.138&93.16&6.91 \\
TN-LN,&$\!\!\!\!\theta\!=\!8.0$&1.033&0.191&0.103&0.039&1.379&2.135&
92.49&6.36 \\ 
GEV&&1.034&0.195&0.106&0.041&1.388&2.134&94.84&8.22 \\
TN-GEV,&$\!\!\!\!\theta\!=\!7.3$&1.033&0.191&0.103&0.039&1.381&2.135&
92.89&6.60 \\\hline 
Ensemble&&1.263&0.211&0.113&0.043&1.441&2.232&45.00&1.80 \\ 
\multicolumn{2}{|l|}{Climatology}&1.550&0.251&0.128&0.045&2.144&2.986&
95.84&11.91 \\ \hline
\end{tabular} 
\caption{Mean CRPS, mean twCRPS for various thresholds \ $r$, \ MAE of
  median and RMSE of mean forecasts and coverage and 
  average width of $96.08\,\%$ central prediction intervals for the
  ECMWF ensemble.} \label{tab:tab2} 
\end{center}
\end{table}

Table \ref{tab:tab2} summarizes the values of various scoring rules
and coverage and average width of $96.08\,\%$ central prediction
intervals. The raw ensemble forecasts outperform the climatological
reference forecast and produce sharp prediction intervals, however, at
the cost of being uncalibrated. All five post-processing methods
significantly improve the calibration and predictive skill of the
ensemble in terms of all scoring rules. The LN and the GEV model show
small improvements over the TN model in terms of the average CRPS, MAE
and RMSE. The best scores are obtained for the TN-LN and TN-GEV
combination models. The TN-LN combination model achieves a minimally
lower MAE and results in slightly narrower central prediction
intervals. Further, the TN, LN and TN-LN models are strictly positive
whereas the GEV and TN-GEV models occasionally assign small non-zero
probabilities to negative wind speed observations. This effect is
typically negligible as the average (maximum) probability mass
assigned to negative wind speeds is smaller than 0.01\,\% (5\,\%) for the
GEV model and smaller than 10$^{-7}$\,\% (0.001\,\%) for the TN-GEV model,
respectively. 

\begin{figure}[b!]
\begin{center}
\includegraphics[width=0.45\textwidth]{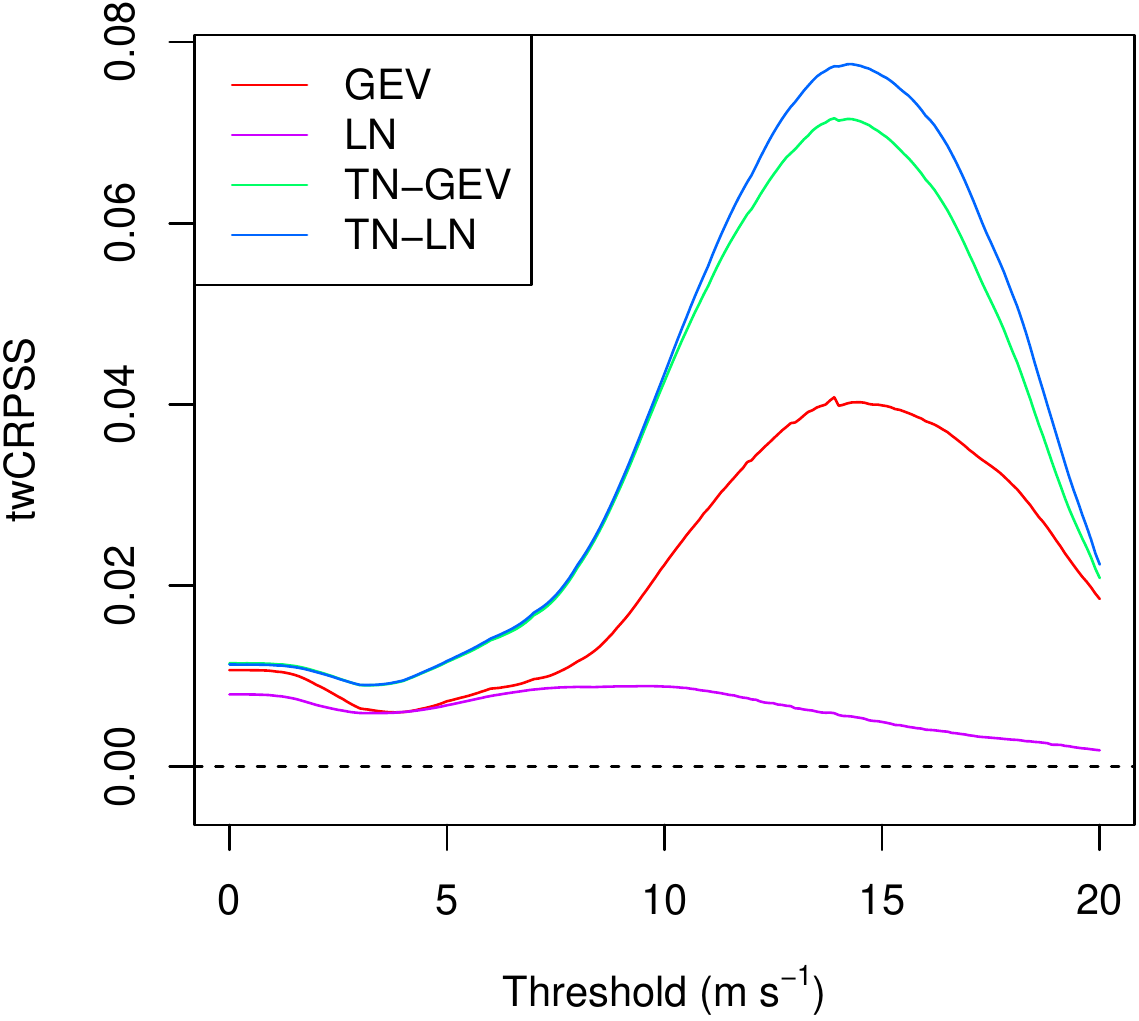}
\caption{twCRPSS values for the ECMWF ensemble with TN as reference model.}   
\label{fig:fig7}
\end{center}
\end{figure} 

To assess the predictive ability for high wind speed observations we also
compute the twCRPS scores at different threshold values, see Table
\ref{tab:tab2}. The  
best scores in the upper tail are obtained by the TN-LN and TN-GEV
combination models and the relative improvements over the TN model are
considerably higher compared to the improvements in the unweighted
CRPS. Figure \ref{fig:fig7} further shows the twCRPSS as a
function of the threshold employed in the indicator weight function
with the TN model as reference forecast. The twCRPSS is strictly
positive for all models and threshold values, indicating improvements
compared to the TN model. Except for the LN model, the twCRPSS of the
models generally increases for larger threshold values and the
greatest relative improvements over the TN model can be detected at
threshold values around $15$ m/s. Despite the decreasing twCRPSS values
of the LN model, the TN-LN model achieves the largest improvements
over the TN model, closely followed by the TN-GEV model. Hence, one
can conclude that the regime-switching models have the best overall
performance showing almost the same predictive skills. 

\begin{figure}[t!]
\subfigure[$\hspace{-1.25cm}$]{\includegraphics[width=0.33
  \textwidth]{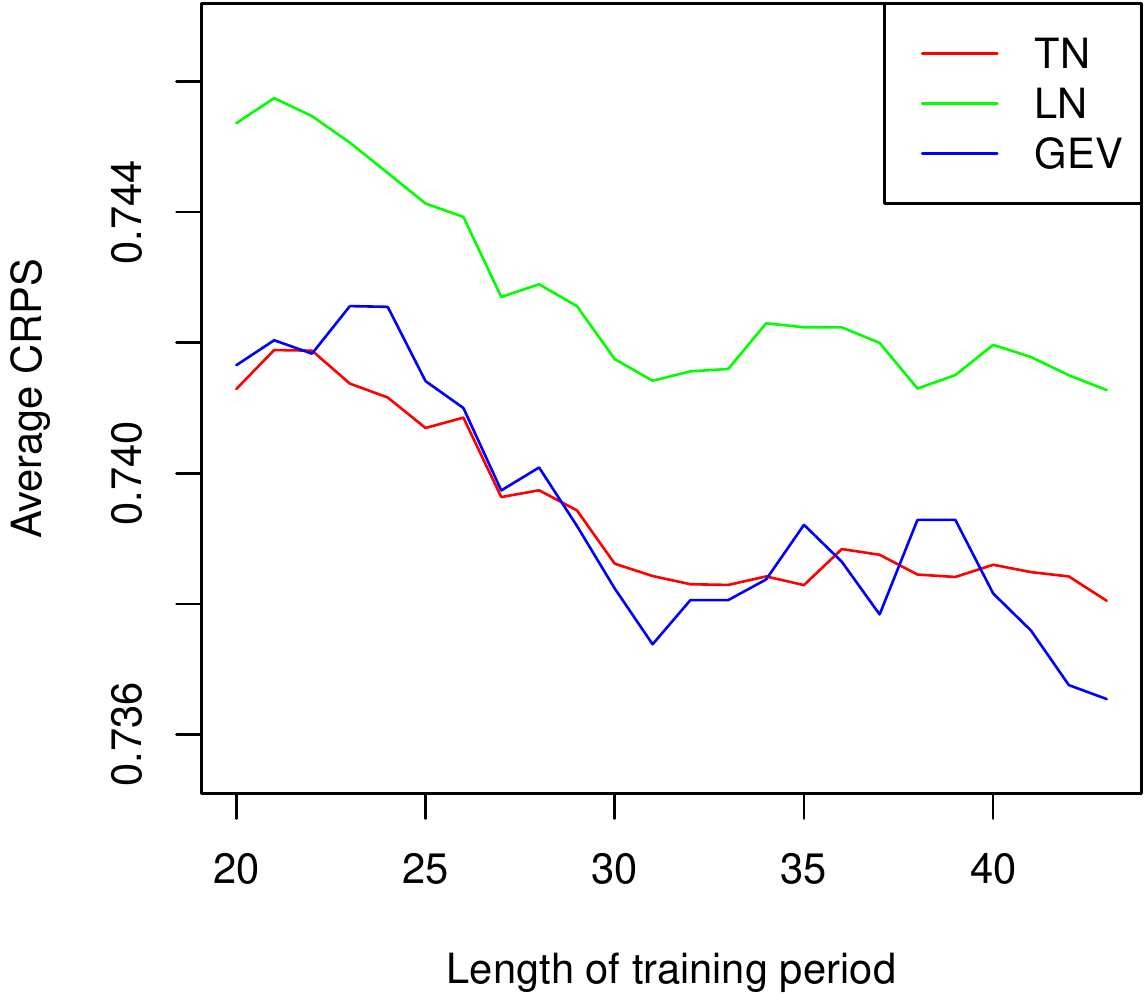}}\hfill  
\subfigure[$\hspace{-1.25cm}$]{\includegraphics[width=0.33
  \textwidth]{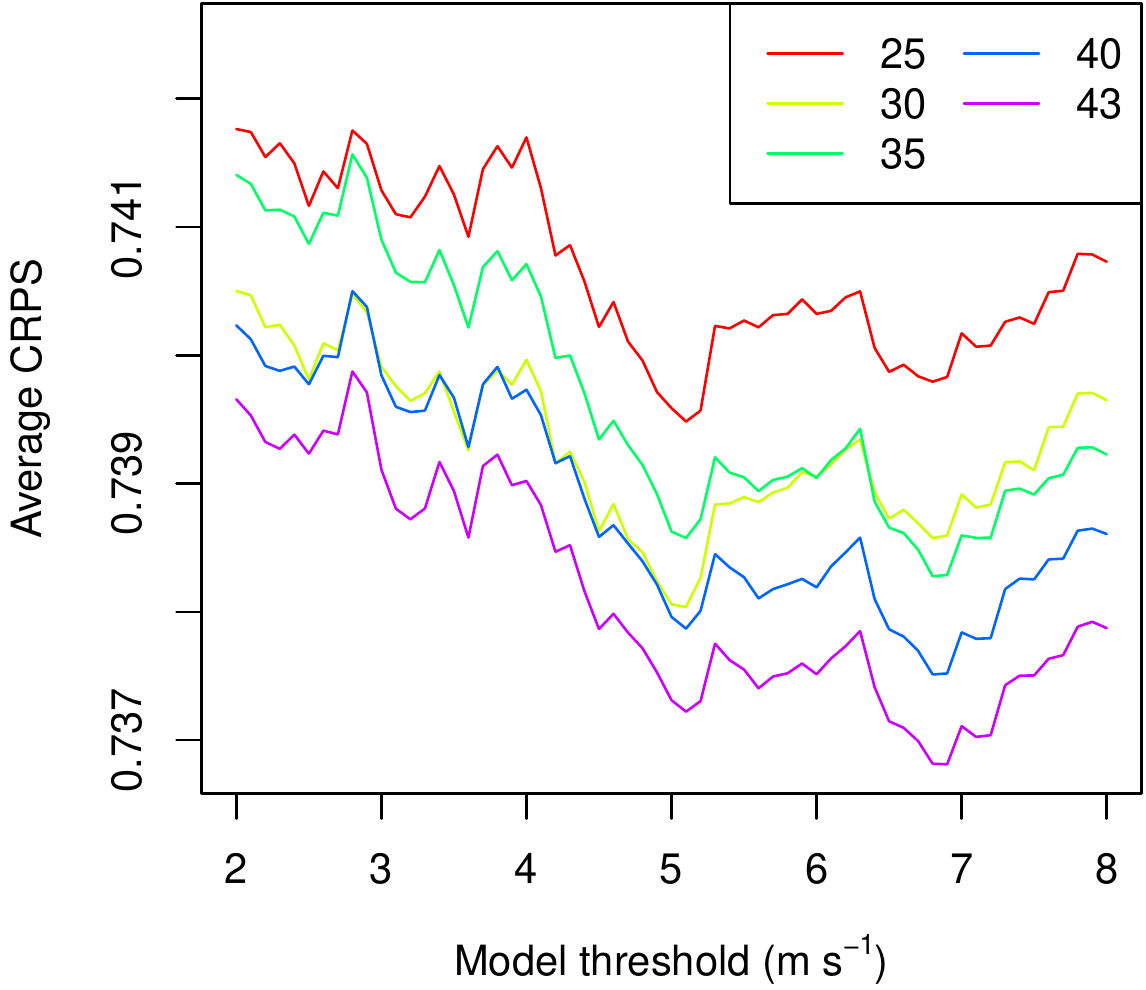}}\hfill 
\subfigure[$\hspace{-1.25cm}$]{\includegraphics[width=0.33
  \textwidth]{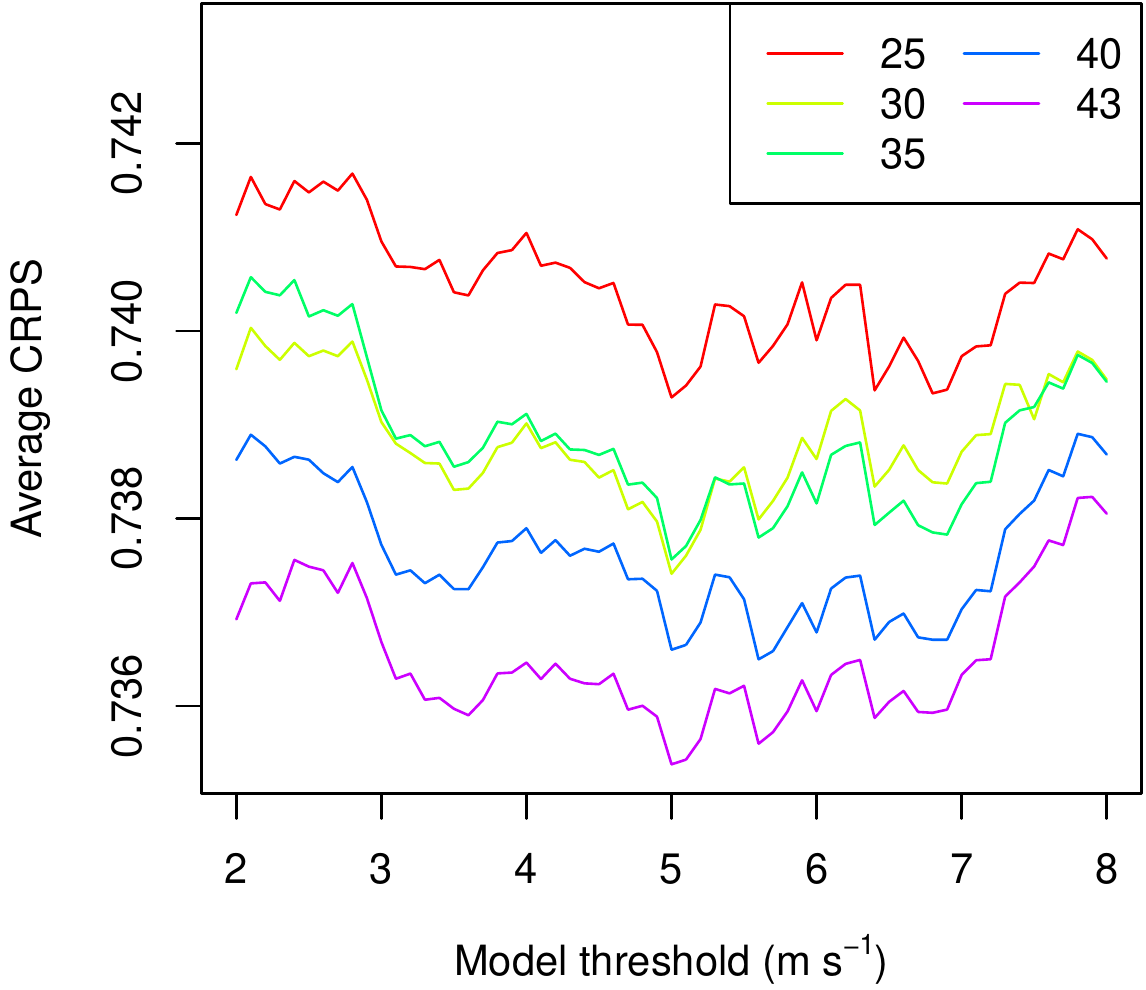}} 
\caption{Mean CRPS values of the (a) EMOS predictive distributions for various 
  training period lengths; (b) TN-LN mixture models corresponding to different
  training period lengths as functions of the threshold; (c) TN-GEV
  mixture models corresponding to different 
  training period lengths as functions of the threshold for the
  ALADIN-HUNEPS ensemble.} 
\label{fig:fig8}
\end{figure}

\subsection{ALADIN-HUNEPS ensemble}
  \label{subs:subs4.3}

The way the ALADIN-HUNEPS ensemble is generated (see Section
\ref{subs:subs2.3}) induces a natural
grouping of ensemble members into two groups. The first
group contains just the control member, while in the second are the 10
statistically indistinguishable ensemble members initialized from
randomly perturbed initial conditions. One should remark here that
in \citet{bhn1} a
different grouping is also suggested (and later investigated in
\citet{bar} and \citet{bhn2}, too), where the odd and even numbered
exchangeable ensemble members form two separate groups. This idea is
justified by the method their initial conditions
are generated, since only five perturbations are calculated and 
then they are added to (odd numbered members) and
subtracted from (even numbered members) the unperturbed
initial conditions. However, since in the present study the results
corresponding to the two- and three-group models are rather similar, only
the two-group case is reported.

\begin{figure}[t!]
\begin{center}
\includegraphics[width=\textwidth]{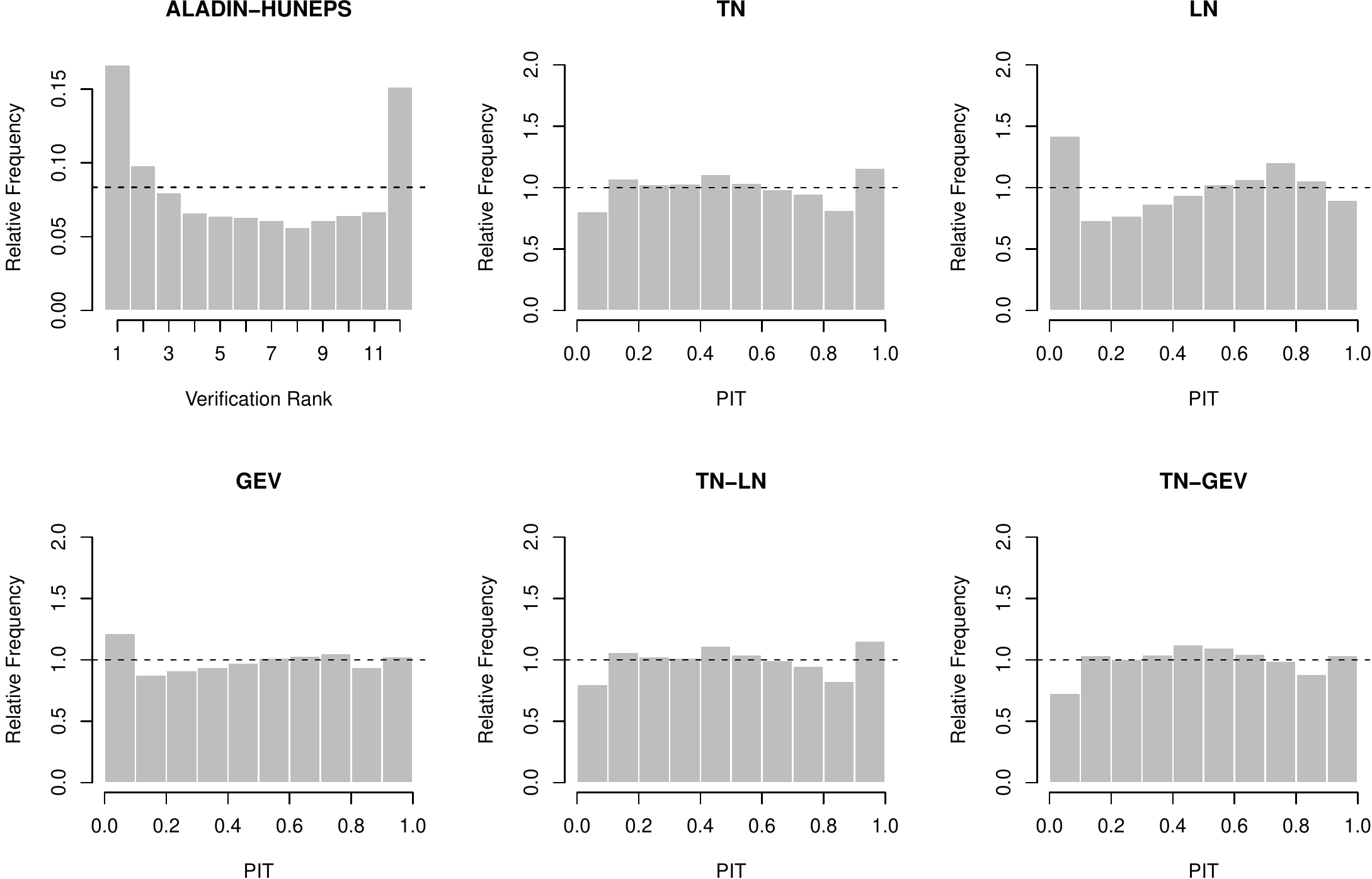}
\caption{Verification rank histogram of the raw ensemble and PIT
  histograms of the EMOS post-processed forecasts  for the
  ALADIN-HUNEPS ensemble.} 
\label{fig:fig9}
\end{center}
\end{figure}  

A detailed earlier study of this particular data set \citep{bhn2}
shows that in case of the TN distribution based EMOS model the optimal length of
the rolling training period for ALADIN-HUNEPS wind speed 
forecasts is 43 days.
Using this training period length one has a verification period
between May 5, 2012 and March 31, 2013
containing 313 calendar days (3\,130 forecast
cases). In order to ensure the comparability of our results to the
earlier studies by having the same verification period we keep the 43
days as the maximum possible training period length. However, based on
Figure \ref{fig:fig8}a showing the mean CRPS values of
TN, LN and GEV models, respectively, as
functions of the length of the training period, this maximal value of 43 days
can also be accepted as optimal for all methods. Furthermore, this
particular length is also supported by Figures
\ref{fig:fig8}b and \ref{fig:fig8}c plotting the mean CRPS values
of the TN-LN and TN-GEV mixture models as
functions of the threshold \ $\theta$ \ for various training period
lengths.
Based on these figures the optimal TN-LN and TN-GEV thresholds are
$6.9$ m/s and $5$ m/s, while the corresponding
percentages of usage of LN and GEV distributions in the mixtures are
$4$\,\% and $15$\,\%, respectively.

\begin{table}[t!]
\begin{center}
\begin{tabular}{|l|c|c|c|c|c|} \hline
EMOS model&TN&LN&TN-LN&GEV&TN-GEV\\\hline
$p$-value&$0.127$&$4.42\!\times\! 10^{-6}$&$0.117$&$0.070$&$0.014$  \\\hline
\end{tabular} 
\caption{$p$-values of Kolmogorov-Smirnov tests for uniformity of PIT
  values for the
  ALADIN-HUNEPS ensemble.} \label{tab:tab3} 
\end{center}
\end{table}

Similar to the previous two sections we first consider the PIT
histograms of all considered EMOS models, displayed in
Figure \ref{fig:fig9}. Compared to
the verification rank histogram of the raw ensemble all
post-processing methods result in a significant improvement in the
goodness of fit to the uniform distribution, while
from the competing calibration methods the TN and the TN-LN mixture
models have the best performance. This latter statement is justified
by the $p$-values of Kolmogorov-Smirnov tests for uniformity given
in Table \ref{tab:tab3}.

\begin{table}[b!]
\begin{center}
\begin{tabular}{|ll|c|c|c|c|c|c|c|c|} \hline
\multicolumn{2}{|l|}{Forecast}&CRPS&\multicolumn{3}{c|}{
  twCRPS $(m/s)$}&MAE&RMSE&Cover.&Av.w.\\\cline{4-6}
\multicolumn{2}{|l|}{}&$(m/s)$&$r\!=\!6$&$r\!=\!7$&$r\!=\!9$&
$(m/s)$&$(m/s)$&$(\%)$&$(m/s)$\\ \hline
TN&&0.738&0.102&0.054&0.012&1.037&1.357&83.59&3.53  \\
LN&&0.741&0.102&0.054&0.011&1.038&1.362&80.44&3.57 \\
TN-LN,&$\!\!\!\!\theta\!=\!6.9$&0.737&0.101&0.054&0.011&1.035&
1.356&83.59&3.54\\ 
GEV&&0.737&0.098&0.052&0.011&1.041&1.355&81.21&3.54 \\
TN-GEV,&$\!\!\!\!\theta\!=\!5.0$&0.735&0.098&0.052&0.011&1.039&
1.355&85.59&3.72 \\\hline 
\multicolumn{2}{|l|}{Ensemble}&0.803&0.112&0.059&0.013&1.069&1.373&68.22&2.88 \\
\multicolumn{2}{|l|}{Climatology}&1.046&0.127&0.064&0.012&1.481&
1.922&82.54&3.43 \\\hline 
\end{tabular} 
\caption{Mean CRPS, mean twCRPS for various thresholds \ $r$, \ MAE of
  median and RMSE of mean forecasts and coverage and 
  average width of $83.33\,\%$ central prediction intervals for the
  ALADIN-HUNEPS ensemble.} \label{tab:tab4} 
\end{center}
\end{table}

\begin{figure}[t]
\begin{center}
\includegraphics[width=0.45\textwidth]{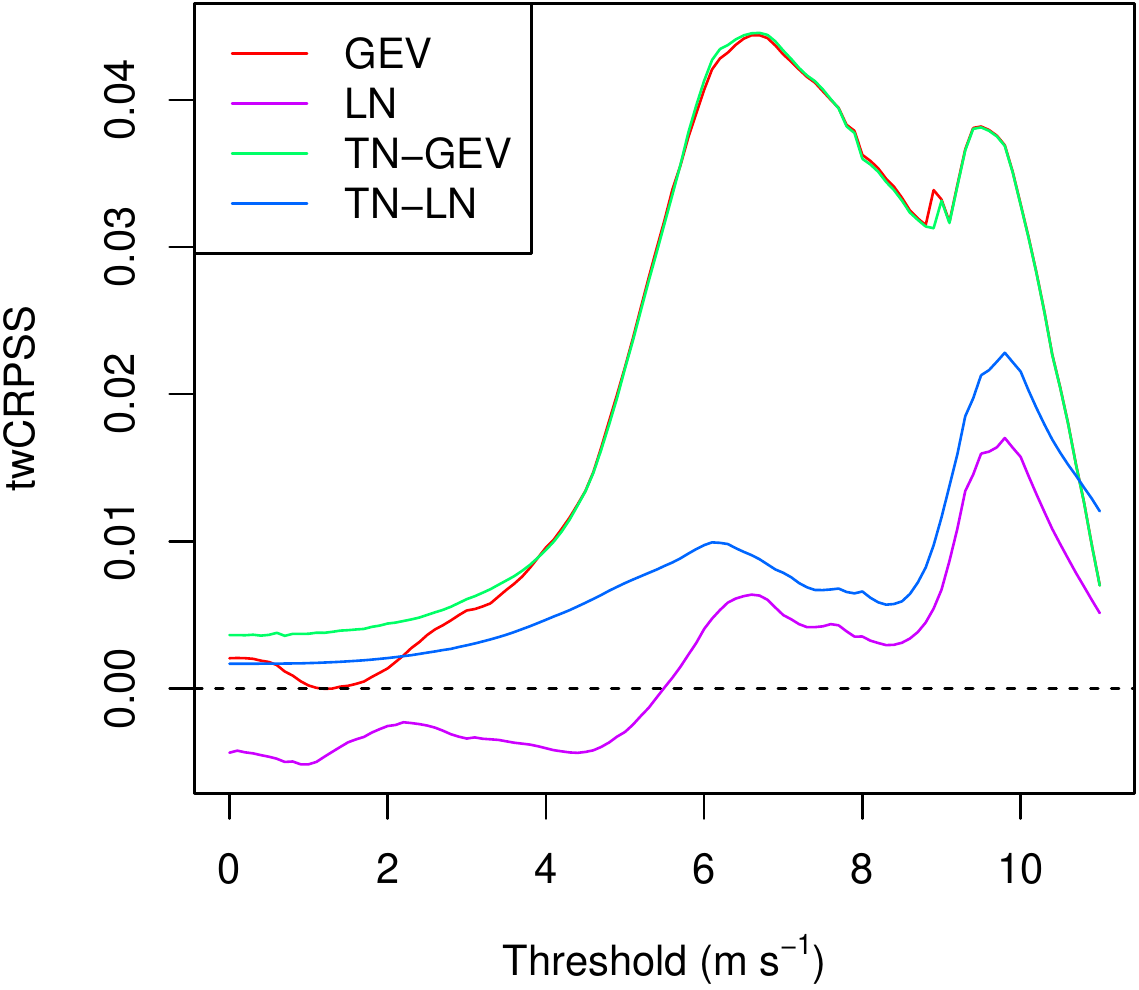}
\caption{twCRPSS values for the ALADIN-HUNEPS ensemble with TN as
  reference model.}    
\label{fig:fig10}
\end{center}
\end{figure}

 Similar to Sections \ref{subs:subs4.1} and \ref{subs:subs4.2} in Table
 \ref{tab:tab4} verification scores for probabilistic 
forecasts and the average width and coverage of $83.33\,\%$ central
prediction intervals are reported. Compared to the raw ensemble
and to climatology post-processed forecasts show the same behaviour as
before: improved predictive skills and better calibration. The lowest
CRPS and RMSE values belong to the TN-GEV mixture model, while the
TN-LN regime-switching method provides the best MAE score and coverage
combined with a rather narrow central prediction interval. Further,
for $6$ m/s 
and $7$ m/s threshold values the GEV and TN-GEV models result in slightly lower
twCRPS scores than the TN-LN mixture, while for $r=9$ m/s this
advantage practically disappears. This phenomenon can also be
observed in Figure \ref{fig:fig10} showing the
twCRPSS values of the GEV, LN, TN-GEV and TN-LN methods with respect to the
reference TN model as functions of the threshold. Finally, the mean
(maximal) probabilities of predicting a negative wind speed by the GEV
and TN-GEV methods are $0.33 \,\%$ ($9.46\,\%$) and $2.74\times 10^{-3}\,\%$
($0.15\,\%$), respectively. Taking also into account the goodness of fit
of PIT histograms (see Table \ref{tab:tab3}) one can conclude that for
ALADIN-HUNEPS wind speed forecasts the TN-LN mixture model has the
best overall performance. 

\section{Conclusions}
  \label{sec:sec5}

We introduce a new EMOS model for calibrating ensemble forecasts of
wind speed providing a predictive PDF which follows a log-normal
distribution. In order to have better forecasts in the tails we also
consider a regime-switching approach based on 
the ensemble median, which considers a truncated normal EMOS model for
low values and a log-normal EMOS for the high ones. The two
approaches are 
tested on wind speed forecasts of the eight-member University of
Washington mesoscale ensemble, of the 
fifty-member ECMWF ensemble and of the eleven-member ALADIN-HUNEPS
ensemble of the Hungarian Meteorological Service. These ensemble
prediction systems differ both in the wind 
speed quantities being forecasted and in the generation of the ensemble
members. Using appropriate verification measures (CRPS of probabilistic,
MAE of median and RMSE of mean forecasts, coverage and average width
of central prediction intervals corresponding to the nominal coverage,
twCRPS corresponding to 90th, 95th and 99th percentiles of the
verifying observations) the predictive performances of the LN and TN-LN mixture
models are compared to those of the TN based EMOS method \citep{tg},
of the GEV and TN-GEV mixture models \citep{lt}, of the raw
ensemble, and of the climatological forecasts as well. From the
results of the presented 
case studies one can conclude that compared to the raw
ensemble and to climatology post-processing always improves the calibration of
probabilistic and accuracy of point forecasts. Further, the
TN-LN mixture model outperforms the traditional TN method \citep{tg}
and it is at least able to keep up with the models utilizing the GEV
distribution \citep{lt} without the problem of forecasting negative wind speed
values.

\bigskip
\noindent
{\bf Acknowledgments.} \  \
Essential part of this work was made during the visit
of S\'andor Baran at the Heidelberg Institute of Theoretical
Studies. Sebastian Lerch gratefully acknowledges support by the
Volkswagen  Foundation within the program
``Mesoscale Weather Extremes -- Theory, Spatial Modelling and
Prediction (WEX-MOP).'' S\'andor Baran was supported by
the Campus Hungary Program and by the T\'AMOP-4.2.2.C-11/1/KONV-2012-0001
project. The project has been supported by the European Union,
co-financed by the European Social Fund.
The authors are indebted to Tilmann
Gneiting for his useful suggestions and
remarks, to the University of Washington MURI group for providing the
UWME data and to Mih\'aly Sz\H ucs from the HMS for the ALADIN-HUNEPS data.

\end{document}